\documentclass[journal=jacsat,manuscript=article]{achemso}

\usepackage{amssymb}
\usepackage{xcolor}
\usepackage{chemformula} 
\usepackage[T1]{fontenc} 

\usepackage[
nonumberlist, 
acronym,      
toc,          
section]      
{glossaries}
\makenoidxglossaries

\newacronym{dft}{DFT}{   Density Functional Theory}
\newacronym{rdm}{RDM}{Reduced Density Matrices}
\newacronym{apg}{APG}{Antisymmetric Product of Geminals}
\newacronym{agp}{AGP}{Antisymmetric Geminal Power}
\newacronym{apsg}{APSG}{Antisymmetric Product of Strongly Orthogonal Geminals}
\newacronym{ap1rog}{AP1roG}{Antisymmetric Product of 1-reference-Orbital Geminals}

\newacronym{apig}{APIG}{Antisymmetric Product of Interacting Geminals}

\newacronym{apr2g}{APr2G}{Antisymmetrized Product of rank-two Geminals}
\newacronym{cc}{CC}{Coupled Cluster}
\newacronym{bcs}{BCS}{ Bardeen–Cooper–Schrieffer}

\newacronym{ci}{CI}{Configuration Interaction}
\newacronym{csf}{CSF}{Configuration State Function}
\newacronym{dmrg}{DMRG}{Density Matrix Renormalization Group}
\newacronym{doci}{DOCI}{Doubly Occupied Configuration Interaction}
\newacronym{fgp}{FGP}{Full Geminal Product}
\newacronym{gvb}{GVB}{Generalized Valence Bond}
\newacronym{gvb-pp}{GVB-PP}{Generalized Valence Bond-Perfect Pairing}
\newacronym{hf}{HF}{Hartree–Fock}
\newacronym{hfb}{HFB}{Hartree–Fock–Bogoliubov}

\newacronym{jagp}{JAGP}{Jastrow Correlated Antisymmetrized Geminal Power}

\newacronym{pt}{PT}{Perturbation Theory}
\newacronym{pccd}{pCCD}{pair-Coupled Cluster Doubles}
\newacronym{qmc}{QMC}{Quantum Monte Carlo}
\newacronym{rg}{RG}{Richardson-Gaudin}

\newacronym{mr-lcc}{MR-LCC}{Linearized Multireference Coupled-Cluster}
\newacronym{russg}{RUSSG}{Restricted Unrestricted SSG}
\newacronym{slg}{SLG}{Strictly Localized Geminals}
\newacronym{ssg}{SSG}{Singlet-type Strongly Orthogonal Geminals}

\newacronym{rpa}{RPA}{Random Phase Approximation}

\author{Pratiksha B. Gaikwad}
\affiliation[Unknown University]
{Department of Chemistry, Quantum Theory Project, University of Florida, Gainesville}
\email{p.gaikwad@ufl.edu}
\author{Krisztina A. Zsigmond}
\affiliation[Unknown University]
{Department of Chemistry, Quantum Theory Project, University of Florida, Gainesville}
\author{Ramón Alain Miranda-Quintana}
\email{quintana@chem.ufl.edu}
\affiliation[Unknown University]
{Department of Chemistry, Quantum Theory Project, University of Florida, Gainesville}

\title[An \textsf{achemso} demo]
  {Geminal Wavefunction Models in Chemistry}

\abbreviations{IR,NMR,UV}

\begin{document}

\begin{tocentry}

\includegraphics[width=0.85\linewidth]{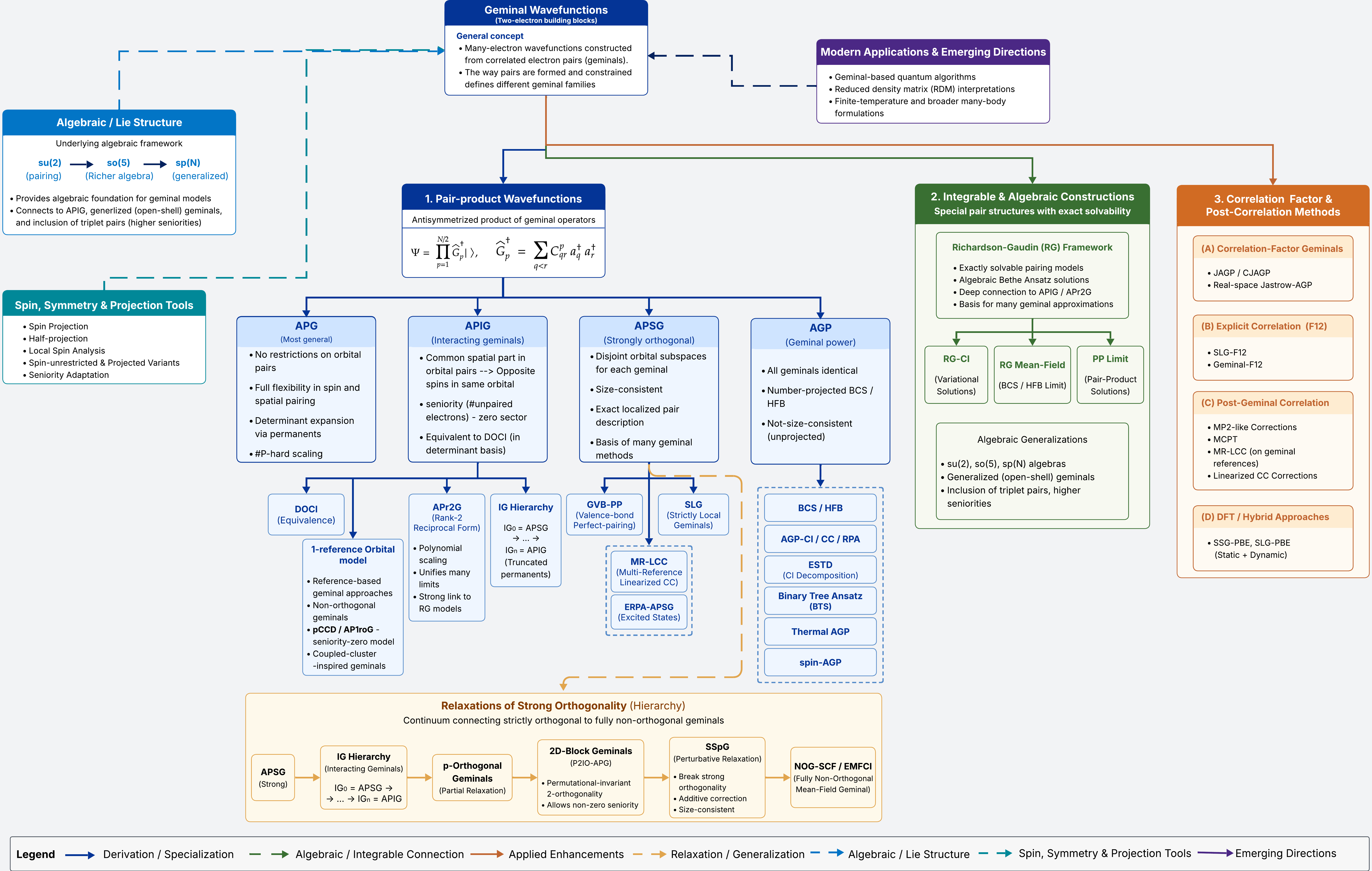}
    





\end{tocentry}


\begin{abstract}

Geminal wavefunctions, introduced in the late 1950s, have long been recognized for their ability to compactly capture strong electron correlation. Despite their promise, they were historically overshadowed by more computationally efficient methods. Advances in both computational resources and theoretical frameworks have renewed interest in geminal-based approaches, particularly as researchers seek accurate yet tractable wavefunctions for complex electronic systems. Recent developments highlight their versatility: from serving as efficient starting points for correlated wavefunctions, to hybrid formulations that blend geminal concepts with coupled-cluster theory, to emerging applications in quantum algorithms where orbital-pairing provides a natural structure. In this mini-review, we summarize key advances in geminal wavefunction theory, with a focus on their modern resurgence, new methodological innovations, and potential directions for electronic structure theory and quantum computation.
\end{abstract}


\section{Introduction}

{\Huge G}eminal wavefunctions provide a compact representation of electron correlation by explicitly pairing electrons, offering a quantum-mechanical realization of the localized bonding picture embodied in Lewis structures. By treating correlated electron pairs as fundamental building blocks, geminal approaches naturally capture key aspects of static correlation that are difficult to describe with single-determinant wavefunctions.

Despite their conceptual simplicity, the practical evaluation of geminal wavefunctions is challenging. The apparent product structure of two-electron building blocks conceals a factorially growing number of orbital-pairing patterns when the wavefunction is expanded in the basis of Slater determinants. As a result, naive implementations quickly become computationally prohibitive. For decades these difficulties limited the broader adoption of geminal-based methods, even though the framework provides an appealing physical interpretation of electron correlation and exhibits deep connections to pairing phenomena in both chemistry and condensed-matter physics.

Historically, the concept of a geminal dates back to Shull’s natural spin-orbital analysis of H$_2$.\cite{shull1959natural} Coleman subsequently placed geminals on a rigorous theoretical footing through his work on $N$-representability and fermionic reduced density matrices.\cite{coleman1963structure} Important developments soon followed, including Arai’s theorem (1965), Kutzelnigg’s antisymmetrized product of strongly orthogonal geminals (APSG, 1967), and the recognition that the antisymmetrized geminal power (AGP) corresponds to a number-projected Hartree–Fock–Bogoliubov (HFB) state. These advances established geminals as a unifying concept linking chemical bonding theory, superconducting pairing models, and quasiparticle descriptions of correlated fermionic systems.

Recent years have witnessed a renewed interest in geminal theories, driven by advances in algorithms, new connections to coupled-cluster and quantum Monte Carlo approaches, and emerging applications in quantum computing. Modern developments have demonstrated that geminal-based wavefunctions can serve both as compact references for strongly correlated systems and as starting points for systematically improvable many-body expansions. A comprehensive overview of recent developments in geminal-based electronic structure methods can be found in the review of Tecmer and Boguslawski.\cite{tecmer2022geminal} In this mini-review, we highlight several representative developments and emerging directions that illustrate the continuing evolution of geminal-based approaches in quantum chemistry.

To provide a coherent perspective, we organize geminal approaches according to their structural principles and physical motivations. We begin by examining families of pair-product wavefunctions, which form the conceptual core of geminal theory. These include general antisymmetrized products of geminals and their important specializations that impose constraints on geminal structure through orbital pairing and their coefficients. We then discuss integrable and algebraic constructions of geminal wavefunctions, focusing on Richardson–Gaudin states and their connections to coupled-cluster and variational many-body approaches. Next, we review correlation-factor geminals, where explicit correlation functions such as Gaussian or Jastrow factors are combined with geminal structures to accelerate basis-set convergence and incorporate short-range electron correlation. Finally, we highlight alternative theoretical perspectives and emerging directions. These include reduced-density-matrix interpretations of geminal correlation as well as recent developments that leverage geminal structures in quantum algorithms for electronic structure. 
Together, these developments illustrate the continuing evolution of geminal-based electronic structure methods and their potential role in future treatments of strongly correlated electrons.

\section{Families of Pair Product Wavefunctions}

Geminal wavefunctions are most naturally classified according to how electron pairs are constructed and how different geminals interact within the many-electron wavefunction. The simplest and historically most important class consists of \textit{pair-product wavefunctions}, in which the many-electron state is written as an antisymmetrized product of two-electron building blocks. Different families of geminal wavefunctions arise by imposing constraints on the structure of these pair operators or on the orbital pairs they contain.

A general pair-product wavefunction for a system of $N$ electrons can written as, 

\begin{equation}
\Psi = \prod_{p}^{N/2} \hat{G}_p^\dagger \, | \rangle, 
\quad 
\hat{G}_p^\dagger = \sum_{q<r} C^{p}_{qr}\, a_q^\dagger a_r^\dagger,
\label{eq:general_geminal}
\end{equation}

where $\hat{G}_p^\dagger$ denotes the creation operator for the $p$-th geminal. Each geminal is expressed as a linear combination of all two-electron creation operators $a_q^\dagger a_r^\dagger$ acting on the vacuum state. 
The operator $a_q^\dagger$ creates an electron in spin orbital $q$, and the coefficients $C^{p}_{qr}$ determine the contribution of each orbital pair $(q,r)$ to the geminal.

In the most general construction, a system of $N$ electrons is described by the $N/2$ geminals, each of which may involve all possible pairs of spin orbitals in the one-electron basis. While the allowed orbital pairs are identical in principle for all geminals, the numerical values of the coefficients $C^p_{qr}$ distinguish one geminal from another. The flexibility of the coefficient tensor therefore determines the expressiveness and computational cost of the resulting wavefunction.

Different families of geminal wavefunctions arise by restricting the allowed structure of the coefficients $C^p_{qr}$ or by imposing constraints on how orbital pairs are distributed among geminals. The most general realization of Eq.~(\ref{eq:general_geminal}) is the \acrfull{apg} wavefunction,  in which each geminal may involve arbitrary combinations of spin-orbital pairs irrespective of spatial and spin parts of the constituting orbitals in the pair. Because the resulting Slater determinant expansion contains permanents of the geminal coefficient matrices, direct evaluation of APG wavefunctions becomes computationally prohibitive for realistic systems.

The antisymmetrized product of geminals (APG) represents the most general pair-product wavefunction, from which several computationally tractable approximations have been derived. Imposing a common orbital basis for all geminals leads to the \acrfull{apig}, while enforcing strong orthogonality between geminals produces the \acrfull{apsg}). When the pairing structure is restricted to opposite-spin electrons occupying the same spatial orbital, the resulting wavefunction belong to the class of \textit{seniority-zero}, where the seniority quantum number counts the number of unpaired electrons and is zero. These constructions form the basis of several widely used wavefunction models including GVB-PP and AP1roG.\cite{calero2025seniority, moisset2022density}

\subsection{Antisymmetrized Product of Geminals - The most general geminal wavefunction}

The antisymmetrized product of geminals (APG) corresponds directly to the general form of Eq.~(\ref{eq:general_geminal}), where each geminal may involve an arbitrary linear combination of spin–orbital pairs. In this formulation the many-electron wavefunction is constructed from $N/2$ independent geminals whose coefficients determine how electron pairs are distributed across the orbital space. Because no restrictions are imposed on the pairing structure, APG provides the most flexible representation within the pair-product framework.

Despite its conceptual simplicity, evaluating APG wavefunctions is computationally demanding. When expanded in the basis of Slater determinants, the coefficients of the determinants are given by matrix permanents of the geminal coefficient matrices. The evaluation of permanents is a $\#P$-hard problem, and the number of pairing patterns grows factorially with system size. Consequently, straightforward implementations of APG quickly become intractable for realistic molecular systems.

\begin{equation} \label{eq:det_expansion}
\begin{split}
| \Psi_{\text{APG}} \rangle &= \prod_{p=1}^{P=N/2} \sum_{q,r}^{2K}  c^{p}_{qr} a_q^\dagger a_r^\dagger | \rangle \\
&= \sum_{\mathbf{m} \in \mathcal{S}} 
\sum_{\{q_1,r_1, ..., q_P, r_P\}=\mathbf{m}}
\text{sgn}(\sigma(q_1,r_1, ..., q_P, r_P))
\begin{vmatrix}
c^1_{q_1,r_1} & \hdots & c^1_{q_P r_P}\\ 
\vdots & \ddots & \vdots \\
c^P_{q_1,r_1} & \hdots & c^P_{q_P r_P}
\end{vmatrix}^{+}
|\mathbf{m} \rangle .
\end{split}
\end{equation}

Here $\mathbf{m}$ denotes a set of spin orbitals occupied in a Slater determinant, and $\mathcal{S}$ is the set of all determinants spanning the Hilbert space. The matrix $\mathbf{C}$ contains the geminal coefficients, whose rows correspond to the $P=N/2$ geminals and whose columns correspond to the pairs of creation operators. The inner sum $\sum_{\{q_1,r_1,\ldots,q_P,r_P\}=\mathbf{m}}$ runs over all possible combinations of creation-operator pairs that generate the determinant $|\mathbf{m}\rangle$. The factor $\text{sgn}(\sigma)$ denotes the signature of the permutation required to reorder the creation operators $a^\dagger_{q_1} a^\dagger_{r_1}\cdots a^\dagger_{q_P} a^\dagger_{r_P}$ into the standard ordering with increasing orbital indices.

The APG ansatz generalizes the number-conserving form of the \acrfull{bcs} superconducting wavefunction, establishing a deep connection between geminal theory and quasiparticle pairing models in condensed-matter physics. However, the expansion of APG in the Slater determinant basis involves matrix permanents, as indicated in Eq.~(\ref{eq:det_expansion}). Because evaluating permanents is a $\#P$-hard problem, the computational cost grows factorially with system size, which historically limited the practical application of APG-based wavefunctions. Consequently, much of the subsequent development in geminal theory has focused on constructing approximations that retain the physical insight of pair-based descriptions while reducing the computational complexity.

Recent algorithmic developments have sought to mitigate this difficulty by identifying the most important pairing structures without enumerating the full combinatorial space. In particular, Richer and co-workers introduced a graphical interpretation of APG overlaps in which pairing configurations are mapped onto weighted matchings of a bipartite graph.\cite{richer2025graphical} Efficient maximum-weight matching algorithms can then identify the dominant pairing patterns, reducing the effective computational scaling from factorial to near-polynomial cost. Methods such as APG-Kps exploit this strategy to recover a substantial fraction of the APG correlation energy while maintaining computational costs comparable to simpler geminal models.

Recent developments have explored low-rank representations of the antisymmetrized product of geminals (APG) as a route to alleviating its formidable computational complexity.\cite{kawasaki2025low} In this approach, the antisymmetric geminal coefficient matrices are decomposed and reconstructed using only their most significant eigenvalues, yielding a hierarchy of low-rank APG wavefunctions that interpolate between compact pair-product models and the full APG limit. Such decompositions exploit the observation that optimized geminal matrices typically possess only a limited number of dominant eigenvalues, allowing the essential electron-pair correlations to be captured with a much smaller parameter space. Building on this idea, Kawasaki and co-workers further addressed the practical challenge of orbital optimization in low-rank APG by introducing a formulation based on direct Givens rotations combined with an error-backpropagation strategy for evaluating gradients.\cite{kawasaki2025efficient} This framework enables analytic optimization of the unitary transformations defining the orbital basis, dramatically reducing the computational cost relative to earlier implementations that relied on numerical differentiation. Benchmark applications to small molecular systems such as H$_2$O and N$_2$ demonstrate that even low-rank variants recover a substantial fraction of the correlation energy and yield potential-energy curves significantly more accurate than AGP or UHF references while maintaining a compact geminal description








\subsection{Antisymmetric Product of Interacting Geminals: Pairing in Spatial Orbitals}

The \acrfull{apig} imposes a common spatial orbital basis for all geminals, such that electron pairs are created in the form $a_{p\alpha}^\dagger a_{p\beta}^\dagger$. This structure leads to a wavefunction closely related to the seniority-zero sector of configuration interaction and is formally equivalent to \acrfull{doci} wavefunction when expanded in the Slater determinant basis.

Although APIG drastically reduces the number of admissible pairing configurations relative to APG, the exact evaluation of its amplitudes still involves matrix permanents, whose computation scales factorially with system size and belongs to the class of $\#P$-hard problems. Consequently, direct implementations of APIG remain computationally demanding despite the physically intuitive pairing structure.

Nevertheless, the APIG ansatz occupies a central position in modern electronic structure theory. By restricting the Hilbert space to seniority-zero determinants, APIG captures the dominant static correlation arising from electron pairing while retaining a compact representation. This structure has motivated the development of several efficient approximations and reformulations, most notably the \acrfull{pccd} or \acrfull{ap1rog} approaches, which reproduce the DOCI wavefunction to high accuracy while scaling comparably to conventional coupled-cluster methods. Beyond quantum chemistry, the APIG framework is also closely connected to pairing Hamiltonians studied in condensed-matter physics and forms a natural bridge to the exactly solvable Richardson–Gaudin models discussed in the following section.

Efforts to mitigate the computational bottleneck of APIG have also led to hierarchical approximations that interpolate between strongly orthogonal and fully interacting geminals. Limacher introduced a wavefunction hierarchy of interacting geminals (IG), which systematically improves the antisymmetrized product of strongly orthogonal geminals (APSG) toward the full APIG limit.\cite{limacher2016new} In this framework the APIG permanent expansion is truncated according to the importance of geminal coefficients, yielding a sequence of approximations (IG$_0$, IG$_1$, IG$_2$, \ldots) where IG$_0$ corresponds to APSG and the highest order recovers APIG exactly. Remarkably, low-order truncations already recover most of the APIG correlation energy while allowing the evaluation of transition density matrices with polynomial scaling. This hierarchy therefore provides a practical route to capturing interactions between electron pairs while avoiding the factorial scaling inherent in the full APIG formulation.

An alternative strategy for constructing tractable interacting geminal wavefunctions was proposed by Johnson \textit{et al.}\cite{johnson2013size} through the \acrfull{apr2g}, shown in eq.(\ref{eq:apr2g}). In this formulation, the geminal coefficients are constrained to the form

\begin{equation}
    c_{i,p} = \frac{1}{a_p \varepsilon_i + b_i \lambda_p}
\end{equation}

which corresponds to the reciprocal of the elements of a rank-two matrix. This structured parametrization is closely related to the algebraic structure of Richardson–Gaudin pairing models. Importantly, it allows the permanent appearing in the Slater determinant expansion of APIG to be evaluated as a ratio of determinants via Borchardt’s theorem, thereby reducing the computational cost from exponential to polynomial scaling.


\begin{equation}\label{eq:apr2g}
\begin{split}
        | \Psi_{\text{APr2G}} (\textbf{a, $\varepsilon$, b $\lambda$}) \rangle &= \prod_{p=1}^{P=N/2} \left( \sum_{i=1}^{K}\frac{a_i^\dagger a_{\bar{i}}^\dagger}{a_p \varepsilon_i + b_i \lambda_p}\right) | \ \rangle, \\ 
        \text{given complex numbers} & \{a_p\}_{p=1}^{P}, \{\varepsilon_i\}_{i=1}^{K}, \{b_i\}_{i=1}^{K}, \{
        \lambda_p\}_{p=1}^{P}
\end{split}
\end{equation}

The APr2G framework unifies several well-known wavefunctions within a single algebraic structure. In particular, appropriate parameter choices recover \acrfull{apg}, APSG, Richardson pairing states, Slater determinants, and number-projected \acrfull{hfb} wavefunctions.\cite{johnson2013size} 

Although the APr2G ansatz provides a tractable approximation to interacting geminal models, it also inherits several limitations common to pair-product wavefunctions. In particular, the basic formulation assumes that electrons are paired and therefore primarily captures intra-pair (static) correlation, while correlations involving unpaired electrons or inter-pair dynamical effects are not fully described. Johnson \textit{et al.} discuss these limitations explicitly and propose several strategies to address them. These include algebraic generalizations of the wavefunction based on larger Lie algebras to incorporate unpaired electrons, as well as projected Schr\"odinger equation approaches for determining the wavefunction parameters. These projected formulations can be solved with polynomial scaling of approximately $\mathcal{O}(K^4-K^5)$ while retaining size-consistency and flexibility associated with non-orthogonal geminals. In this way, the APr2G framework serves both as a computationally tractable model for strong correlation and as a foundation for more general geminal-based wavefunction constructions.

\subsection{Antisymmetric Product of Strongly Orthogonal Geminals}

In contrast to APIG, which allows different geminals to share orbitals, \acrfull{apsg} imposes the condition of \textit{strong orthogonality}, meaning that the orbital spaces associated with different geminals are mutually exclusive. As a consequence, each spatial orbital belongs to at most one geminal, and the total wavefunction becomes an antisymmetrized product of independent electron pairs occupying disjoint orbital subspaces. Historically, APSG emerged as one of the first practical geminal wavefunctions capable of describing localized electron pairs within a fully variational framework.

The concept of orthogonality plays a central role in geminal theory. One may distinguish between \textit{weak orthogonality}, which requires orthogonality of geminals as two-electron functions, and the stronger condition of \textit{strong orthogonality}, in which the one-electron orbital spaces associated with different geminals are mutually exclusive. The latter constraint greatly simplifies the evaluation of Hamiltonian matrix elements and forms the basis of the APSG approximation.

In second-quantized form, the APSG wavefunction can be written as

\begin{equation}
        \Psi_{\text{APSG}} = \prod_i \psi_i^\dagger | \ \rangle = \psi_1^\dagger \psi_2^\dagger \hdots \psi_{P}^\dagger | \rangle, \qquad P= N/2
\end{equation}

where each operator $\psi_i^\dagger$ creates an electron pair belonging to geminal $i$,

\begin{equation}
    \psi_i^\dagger = \sum_{p,q \in i} c_{pq}^{(i)} a_{p}^\dagger a_{q}^\dagger 
\end{equation}

Here, $p$ and $q$ denote spin orbtals belonging to the orbital subspace associated with geminal $i$. The restriction $p,q \in i$ indicates that each gemianl is expanded only within its own orbital subspace. 

Mathematically, the strong orthogonality condition requires that any two different geminals occupy orthogonal one-electron subspaces. In the first-quantized representation this can be expressed as

\begin{equation}
\int \psi_i(\mathbf{r}_1,\mathbf{r}_2)\,
\psi_j(\mathbf{r}_1,\mathbf{r}_2)\,
d\mathbf{r}_1 = 0,
\qquad i \neq j ,
\end{equation}

which ensures that the two-electron functions are orthogonal with respect to the coordinates of either electron. Arai’s theorem shows that this condition is equivalent to expanding each geminal in a set of mutually exclusive orthogonal orbital subspaces.\cite{arai1960theorem}
As a consequence, the first-order reduced density matrix $P_{pq}$ becomes block-diagonal with respect to geminal subspaces (eq.(\ref{eq:block-diag})), which greatly simplifies the evaluation of matrix elements and contributes to the favorable computational scaling of APSG.
\begin{equation} \label{eq:block-diag}
    P_{pq} = \langle\Psi | a_q^\dagger a_p | \Psi \rangle = \delta_{ik} P_{pq}^{(i)}, \qquad \text{for} \ p \in i, q\in k.
\end{equation}

The APSG constitutes the canonical pair-product wavefunction in which each electron pair occupies a mutually exclusive orbital subspace. Its formal structure, density matrices, and algebraic properties have been discussed extensively in earlier reviews.\cite{surjan1999introduction,surjan2012strongly} By enforcing strong orthogonality, APSG achieves size-extensivity, a variational upper bound to the exact energy, and a natural exponential representation that connects it formally to a restricted coupled-cluster doubles ansatz. 

Chemically, this structure provides a localized description of bonding: each geminal spans a two-electron subspace that often corresponds to a Lewis-type bond. As a result, APSG yields smooth potential energy surfaces and robust descriptions of single bond dissociation. However, the strong orthogonality constraint eliminates inter-geminal correlation, preventing the description of dispersive interactions, collective delocalization, and certain excited states. These limitations motivated the development of systematic correction strategies.

Several early improvement schemes were proposed, including Geminal-CI expansions, multiconfigurational perturbation theories (MCPT), Dyall-type perturbative corrections, and constant-denominator approaches.\cite{surjan2012strongly} Although these 
methods improved total energies and size-consistency relative 
to CISD, they did not establish APSG as a universally reliable 
correlation model.

The extension of APSG to periodic systems illustrates its conceptual flexibility. Tokmachev and Dronskowski developed a variational APSG formulation for polymers and one-dimensional solids under periodic boundary conditions.\cite{tokmachev2006geminal} By constructing the wavefunction as an antisymmetrized product of local singlet geminals assigned to each unit cell, they achieved Hartree-Fock (HF)–like computational scaling while incorporating explicit pair correlation. Applications to chains of hydrogen, LiH, and carbon showed that the geminal ansatz captures bond alternation, Peierls distortions, and correlation-driven structural instabilities in extended systems.

Despite its appealing locality, purely singlet-coupled APSG can fail to reproduce correct fragment spin behavior in multiple bond dissociation. Because only singlet pairs are included, the dissociation limit may contain an unphysical admixture of spin states, leading to incorrect local spin values on the fragments. 

One way to remedy this deficiency is to allow singlet–triplet mixing
within the geminal. In this case the $i$-th geminal can be written as a linear combination of singlet and triplet pair functions

\begin{equation}
\psi_i = \sigma_i\, {}^{1}\psi_i + \tau_i\, {}^{3}\psi_i ,
\end{equation}

with normalization condition $\sigma_i^2 + \tau_i^2 = 1$. Such mixed-spin geminals permit a more flexible description of fragment spin states while retaining the pair-product structure of the APSG wavefunction.\cite{jeszenszki2015local}

Analysis of local spin expectation values provides a useful diagnostic of this limitation, revealing that singlet-coupled geminals underestimate high-spin character of dissociated fragments.\cite{jeszenszki2015local} Allowing singlet–triplet mixing at the geminal level restores the correct asymptotic spin behavior, demonstrating that triplet components are qualitatively necessary for bond breaking within strongly orthogonal frameworks.

Beyond its intrinsic valence-bond character, APSG has increasingly been recognized as a compact multireference reference state for perturbative treatments.\cite{jeszenszki2014perspectives} Because each geminal spans an exact two-electron subspace while inter-geminal excitations are absent, APSG occupies an intermediate position between complete active space description and incomplete model space lacking configurations that transfer electrons between geminals. This structure naturally motivates perturbative corrections that incorporate inter-geminal excitations while preserving the variational static reference.

Subsequent developments followed two complementary strategies: (i) augmenting APSG with systematic dynamical correlation corrections, and (ii) relaxing the structural constraints of strong orthogonality while retaining polynomial scaling.

Several approaches illustrate these directions. One example is antisymmetrized product of the Singlet-type Strongly Orthogonal Geminals (SSG) with perturbatively correcttions, together referred as SS$_p$G. In this approach, dynamical correlation is incorporated by relaxing the strong orthogonality at the perturbative level, yielding size-consistent corrections with favorable scaling and improved accuracy relative to strictly orthogonal models.\cite{cagg2014sspg} Alternatively, hybrid static–dynamic frameworks combine geminal references with density functionals, as in the \acrfull{ssg} (PBE) model, where a scaled correlation functional supplements the static geminal energy.\cite{cagg2012density}

Explicitly correlated corrections offer another systematic route to include dynamic correlation. The SLG–F12 framework employs strictly localized geminals (SLG) as a static reference and incorporates all double excitations within an internally contracted F12 correction.\cite{mihalka2025combining} Because the strong orthogonality partitions the orbital space into independent geminal subspaces, the resulting operators and density matrices acquire a block structure (eq. (\ref{eq:block-diag})) that simplifies evaluation of F12 terms. Benchmark calculations on small biradicals (O$_2$, NH, CH$_2$) show that the most complete SLG–F12 variants recover about 85–90\% of the SLG-PT2 correlation energy and yield singlet–triplet gaps typically within $\sim$0.01–0.1 eV of experimental values while exhibiting accelerated basis-set convergence.

Finally, group-function generalizations extend strictly local geminals beyond simple two-electron bonds. In the SLG/SCF framework, strictly local geminals are combined with conventional SCF descriptions of selected electron groups, allowing radicals and delocalized $\pi$ systems to be treated variationally while retaining near-linear scaling for local subsystems.\cite{tokmachev2006group} This work reframed strongly orthogonal geminals not as a purely localized bond model, but as a flexible component within hybrid multireference architectures.


A more systematic treatment of dynamical correlation was achieved through a \acrfull{mr-lcc} correction built on the APSG reference\cite{zoboki2013linearized}. Unlike second-order perturbation theory, which suffers from zero-denominator divergences when geminals become degenerate, the MR-LCC formulation derives from a truncated exponential ansatz and yields a uniquely solvable, size-extensive set of linear amplitude equations. Excitations are expressed in terms of geminal creation and annihilation operators, enabling a unified description of dispersive, charge-transfer, and spin-polarized effects. Applications to bond dissociation and weak intermolecular interactions demonstrated qualitatively correct potential energy curves and near-FCI accuracy for dispersion-dominated systems. These results positioned APSG as a viable coupled-cluster–type reference rather than merely a perturbative starting point.

A formally grounded route to restoring intergeminal dynamical correlation was introduced by Pernal, who derived an intergeminal correction from a fluctuation–dissipation framework expressed through ERPA or TD-APSG transition density matrices.\cite{pernal2014intergeminal} The resulting ERPA–APSG functional preserves size consistency, remains exact for two-electron systems, and reduces asymptotically to the correct second-order dispersion interaction for weakly interacting subsystems. Numerical applications to He$_2$, Li$_2$, H$_2$O, FH, and hydrogen-bonded dimers demonstrate recovery of both short- and long-range correlation while maintaining the qualitatively correct APSG dissociation behavior.


\subsubsection{Group-Function and Geminal Mean-Field Methods}

A systematic framework for relaxing the strong orthogonality constraint in geminal-based wavefunctions was developed by Cassam-Chena\"i and co-workers through the electronic mean-field configuration interaction (EMFCI) methodology.\cite{cassam2006electronic, cassam2007electronic, cassam2010electronic} In this approach, the many-electron wavefunction is expressed as an antisymmetrized product of electron-group functions, and the Schr\"odinger equation is solved iteratively by optimizing one group in the mean field generated by the others. The resulting procedure resembles a configuration interaction calculation performed within contracted electron groups, followed by self-consistent optimization of the group functions. Unlike traditional group-function or APSG formulations, EMFCI does not impose strong orthogonality between geminals, allowing inter-group correlation to be treated variationally.

Subsequent developments introduced practical algorithms and integral formulas enabling efficient evaluation of Hamiltonian and overlap matrix elements for general group-function wavefunctions, exploiting algebraic structures such as Hopf algebras to derive recursive expressions for matrix elements.\cite{cassam2006electronic} Applications to small atoms and molecules demonstrated that the resulting nonorthogonal geminal self-consistent field (NOG-SCF) scheme can recover a large fraction of the correlation energy while maintaining a compact wavefunction representation. Together, these works established a flexible hierarchy of geminal mean-field methods that interpolate between strongly orthogonal geminal models and more general interacting geminal descriptions.

\subsubsection{2D-Block Geminals: Controlled Relaxation of Orthogonality}

While strong orthogonality greatly simplifies the evaluation of matrix elements in APSG, it also suppresses inter-geminal correlations and restricts the wavefunction largely to seniority-zero configurations. A natural strategy to recover additional variational flexibility is therefore to relax the orthogonality constraints imposed between geminals.

Cassam-Chena\"i and co-workers introduced a hierarchy of APG models based on the concept of \textit{$p$-orthogonality}, which generalizes the notion of strong orthogonality.\cite{cassam20232d} Two many-electron wavefunctions are said to be $p$-orthogonal when their $p$-particle internal Hilbert spaces are orthogonal. In this hierarchy, $p=1$ corresponds to strong orthogonality (as in APSG), while increasing $p$ progressively relaxes the constraint. In particular, \textit{2-orthogonality} allows different geminals to share orbital subspaces provided that their two-electron components remain orthogonal. This weaker condition preserves electron indistinguishability while enabling additional inter-geminal correlations.

Building on this idea, Cassam-Chena\"i \textit{et al.} proposed the \textit{permutationally invariant 2-orthogonality} (PI2O) constraints for APG wavefunctions.\cite{cassam20232d} When expressed in the matrix representation of geminals,

\begin{equation}
\Phi_k =
\sum_{i,j} (C_k)_{ij}\,\phi_i \wedge \phi_j ,
\end{equation}

these constraints translate into algebraic conditions on the geminal matrices involving traces of their products. A particularly convenient class of solutions is obtained when the matrices $C_k$ adopt a block-diagonal structure,

\begin{equation}
C_k = \mathrm{diag}\left(B^{(k)}_1, B^{(k)}_2, \ldots \right),
\qquad
B^{(k)}_l \in \mathbb{C}^{2\times2},
\end{equation}

where each block is drawn from a restricted set of $2\times2$ matrices (e.g., Pauli matrices or parametrized diagonal forms). This construction partitions the orbital space into subspaces of dimension at most two and leads to the so-called \textit{2D-block geminal} ansatz.

The resulting block structure dramatically reduces the number of trace terms appearing in the general APG overlap formula, thereby restoring computational tractability while allowing configurations of mixed seniority and partial orbital overlap between geminals.\cite{cassam20232d} Proof-of-principle calculations show that the 2D-block model yields variational energies systematically lower than APSG for small molecules while retaining polynomial computational scaling.

A subsequent study provided practical guidelines for selecting effective orbital partitions and excitations within this framework.\cite{cassam20252d} By starting from APSG-optimized Arai orbital spaces and grouping orbitals according to natural occupation numbers, the combinatorial growth of block assignments can be significantly reduced without sacrificing variational flexibility. Applications to small diatomics demonstrate improvements over APSG in both equilibrium energies and spectroscopic properties while maintaining chemically interpretable pair structures.

Together, these developments position 2D-block geminals as a controlled relaxation of APSG, occupying an intermediate regime between strictly orthogonal pair models and fully non-orthogonal APG wavefunctions. By enlarging the local block structure rather than abandoning orthogonality altogether, the approach incorporates inter-geminal and non-seniority-zero correlation while retaining polynomial complexity and chemical interpretability.

Overall, APSG and its refinements illustrate a recurring theme in geminal theory: strong orthogonality provides a compact and chemically transparent static-correlation backbone, but systematic treatment of inter-geminal coupling, whether perturbative, Coupled Cluster–based, explicitly correlated, or block-relaxed, remains essential for quantitative accuracy.

\subsubsection{Algebraic Generalizations of Geminal Wavefunctions}

A recent algebraic generalization of geminal wavefunctions that incorporates the open-shell singlet pairs was developed by Johnson \textit{et al.} based on the Lie-algebraic structure of electron pairs\cite{johnson2025singlet}.
Conventional geminal models such as APSG and APIG typically restrict electron pairs to closed-shell singlets occupying the same spatial orbital (seniority-zero). Johnson \textit{et al.} construct operators that create singlet pairs distributed across different orbitals, thereby enabling explicit treatment of open-shell singlet correlations.

For example, closed-shell pair operators acting on spatial orbital $i$ can be written as
\begin{equation}
S_i^{+}=a_{i\uparrow}^{\dagger}a_{i\downarrow}^{\dagger}, \quad
S_i^{-}=a_{i\downarrow}a_{i\uparrow}, \quad
S_i^{z}=\tfrac12\left(a_{i\uparrow}^{\dagger}a_{i\uparrow}+a_{i\downarrow}^{\dagger}a_{i\downarrow}-1\right),
\end{equation}
which respectively create, annihilate, and count closed-shell electron pairs.

Singlet pairs spanning different orbitals are created by
\begin{equation}
A_{ij}^{+}=\tfrac12\left(a_{i\uparrow}^{\dagger}a_{j\downarrow}^{\dagger}-a_{i\downarrow}^{\dagger}a_{j\uparrow}^{\dagger}\right).
\end{equation}

A general singlet geminal operator can then be written as
\begin{equation}
    G^{\dagger}=\sum_{ij} g_{ij} A_{ij}^{+},
\end{equation}

from which many-electron wavefunctions are constructed as products of such geminals.  
Different geminal ans\"atze can be interpreted as particular restrictions on the coefficients $g_{ij}$, as summarized in Table~\ref{tab:geminal_constraints}.

\begin{table}[h]
\centering
\caption{Representative restrictions on the general singlet geminal operator 
$G^\dagger=\sum_{ij} g_{ij}A_{ij}^+$ leading to commonly used geminal wavefunctions.}
\label{tab:geminal_constraints}
\begin{tabular}{lll}
\hline
Constraint on $g_{ij}$ & Resulting operator form & Corresponding model \\
\hline
$g_{ij}=g_i\delta_{ij}$ &
$G^\dagger=\sum_i g_i S_i^+$ &
APIG \\

$G_\alpha^\dagger$ acts on disjoint orbital subsets &
$|\Psi\rangle=\prod_\alpha G_\alpha^\dagger|0\rangle$ &
APSG \\

$G_\alpha^\dagger = G^\dagger$ for all $\alpha$ &
$|\Psi\rangle=(G^\dagger)^{N/2}|0\rangle$ &
AGP \\

$g_{ij}$ nonzero only within orbital blocks &
Block-local pair operators &
2D-block geminals \\
\hline
\end{tabular}
\end{table}

Within this formalism, the algebraic structure of the pair operators emerges naturally. Each spatial orbital defines a set of operators obeying the $\mathrm{su}(2)$ algebra that describe creation, annihilation, and counting of closed-shell pairs. When singlet pairs spanning two orbitals are included, the operator set expands to an $\mathrm{so}(5)$ algebra. Extending this construction to $N$ spatial orbitals yields the symplectic algebra $\mathrm{sp(N)}$, which generates the full space of singlet pair operators.

The resulting $\mathrm{sp}(N)$ geminal product admits a projected Schr\"odinger equation (pSE) formulation in a \acrfull{csf} basis, composed of spin-adapted linear combinations of Slater determinants. In this representation the expansion coefficients take the form of mixed discriminants (generalized combinatorial sums related to matrix permanents), while the overlaps among basis states are governed by combinatorial structures related to cosets of the symmetric group, reflecting the permutation symmetries of electron pairs.

Although the unrestricted $\mathrm{sp}(N)$ formulation is generally computationally intractable, the algebraic analysis clarifies how tractable models arise from particular structures imposed on the geminal coefficients. For example, factorized coefficients reduce the wavefunction to a Slater determinant, symmetric coefficients recover \acrfull{agp}-like forms, and block-restricted constructions connect to approaches such as 2D-block geminals. These observations help place established geminal ans\"atze—including APSG and related pair-product models—within a broader algebraic framework.

Numerical tests on small strongly correlated systems demonstrate that allowing open-shell singlet pairs improves upon purely closed-shell geminal descriptions, indicating that seniority-zero restrictions alone are insufficient for capturing bond-breaking and related strong-correlation effects.

Overall, this work situates traditional geminal wavefunctions within a broader Lie-algebraic hierarchy of singlet pair constructions and highlights possible directions for systematically extending tractable pair-product models beyond seniority-zero approximations.


\subsection{Antisymmetric Geminal Power}

The \acrfull{agp} occupies a central position among geminal wavefunctions and represents the simplest global pairing ansatz beyond independent-pair models. In contrast to the general pair-product form of Eq.(\ref{eq:general_geminal}), where distinct geminals are assigned to each electron pair, the AGP wavefunction is constructed by repeatedly applying a single collective geminal operator,

\begin{equation} \label{eq:agp}
|\Psi_{\mathrm{AGP}}\rangle
=
\left(\hat{G}^\dagger\right)^{N/2} |0\rangle ,
\qquad
\hat{G}^\dagger =
\sum_{q<r} C_{qr}\, a_q^\dagger a_r^\dagger .
\end{equation}

Thus, all electron pairs are drawn from the same geminal, allowing electrons to occupy different orbital pairs in a fully delocalized manner. In its most familiar form, AGP can also be interpreted as the number-projected \acrfull{hfb} or BCS state expressed in the natural orbital basis, providing a compact description of collective pairing correlations while remaining within a seniority-zero Hilbert space.

Recent developments have explored alternative ways of organizing many-body wavefunctions around the AGP structure. Uemura, Kasamatsu, and Sugino proposed a representation of configuration-interaction (CI) wavefunctions based on a canonical tensor decomposition of the CI coefficient tensor.\cite{uemura2015configuration} In their extended symmetric tensor decomposition (ESTD) formulation, the CI expansion is reorganized into a linear combination of antisymmetrized geminal powers, effectively expressing the full-CI wavefunction as a compact series of AGP states with rapidly convergent total energies.

Considering that HF is often not a qualitatively correct reference in the presence of strong correlation, the usefulness of AGP as a reference for correlated methods—\acrfull{ci}, \acrfull{cc}, and \acrfull{rpa}—has been actively investigated. Henderson and Scuseria examined post-AGP correlation strategies, including AGP-CI\cite{henderson2019geminal}, AGP-LCC, and AGP-RPA, within a “project-then-correlate’’ framework in which a symmetry-projected AGP serves as the reference state for subsequent correlation treatments\cite{henderson2020correlating}. The authors study the reduced BCS Hamiltonian, where pairing correlations dominate and conventional single-reference approaches fail dramatically. 

To describe pairing correlations in this model, the formalism is naturally expressed in terms of pair operators and the AGP wave function, which constitute the central working equations of the approach:

\begin{equation}\label{eq:operators}
    \begin{split}
        \text{Pair creation operator} & \quad P_p^\dagger = a_{p}^\dagger a^\dagger_{\bar{p}} \\
        \text{Pair annihilation operator} & \quad P_p = a_{p} a_{\bar{p}}        \\
         \text{Number operator} & \quad N_p = a^\dagger_p a_p + a^\dagger_{\bar{p}} a_{\bar{p}} \\
         \text{AGP Wavefunction} & \quad \Psi_{\text{AGP}} = \frac{1}{P!} (\Gamma)^P | \ \rangle  = \sum_{p_1 < \hdots < p_n} \eta_{p_1} \hdots \eta_{p_n} P^\dagger_{p_1} \hdots P^\dagger_{p_n} | \ \rangle \\
         \text{AGP Killing Operator} & \quad K_{pq} = \eta_p^2 P_p^\dagger P_q + \eta_q^2 P_q^\dagger P_p  + \frac{1}{2} \eta_p \eta_q (N_pN_q - N_p - N_q)
    \end{split}
\end{equation}

Because AGP already captures essential pairing correlations, these post-AGP methods are primarily tasked with recovering the remaining residual correlation. In this context, correlation is introduced by replacing Hartree–Fock excitation operators with operators derived from the AGP killing operators, ensuring a consistent description within the geminal framework. However, as the AGP reference approaches the Hartree–Fock limit, numerical instabilities arise due to singularities in the metric associated with AGP excitation operators, whose norms vanish in this limit. These issues complicate iterative implementations of CI- and CC-like approaches, particularly because the excitation manifold becomes overcomplete and ill-conditioned.

Among the proposed extensions, AGP-RPA appears particularly promising because it avoids explicit construction of high-rank excitation spaces and instead relies on commutator-based equations of motion, requiring only low-order reduced density matrices and exhibiting favorable computational scaling.

Several works have explored structural generalizations of AGP that increase its variational flexibility while retaining polynomial scaling. A recent example is the binary tree state (BTS) ansatz introduced by Dutta \textit{et al.}\cite{dutta2024correlated} BTS preserves the monomial character of AGP coefficients while lifting the complete symmetry constraint inherent to elementary symmetric polynomials. 

In the BTS formulation, the wavefunction in the seniority-zero sector can be written as

\begin{equation}
    |Bn_m \rangle = \sum_{1 \leq p_1 < \hdots < p_n \leq m} \eta^1_{p_1} \eta^2_{p_2} \hdots \eta^n_{p_n} P^\dagger_{p_1} \hdots P^\dagger_{p_n} | \ \rangle
\end{equation}

where $P_p^\dagger$ is the pair creation operator mentioned in eq.(\ref{eq:operators}), and the coefficients $\eta^j_p$ define the pair amplitudes associated with each level of the binary-tree construction. 

Formally, BTS approximates the permanent structure of the antisymmetrized product of interacting geminals (APIG) by retaining a single monomial contribution from each permanent expansion, resulting in a polynomial-cost evaluation based on binary tree recursions. Unlike AGP, the BTS ansatz is rigorously size-consistent, as tensor products of subsystem BTS states yield a BTS for the composite system. Benchmark calculations on pairing and spin Hamiltonians demonstrate systematic improvement over AGP while maintaining manageable computational scaling. Moreover, BTS can be further correlated via Hilbert-space Jastrow coupled-cluster corrections or linear combinations of binary tree states, establishing it as an intermediate level in a hierarchy connecting AGP to fully interacting geminal models.

Beyond traditional electronic-structure applications, AGP has recently been generalized to other quantum many-body systems. Liu \textit{et al.}\cite{liu2023exploring} reformulated AGP in terms of the spin $\mathfrak{su}(2)$ algebra, enabling its application to lattice spin Hamiltonians such as the XXZ and $J_1$-$J_2$ Heisenberg models. By exploiting the algebraic correspondence between fermionic pairing operators and spin ladder operators, they introduced a spin-AGP (sAGP) wavefunction defined as

\begin{equation}
    |\mathrm{sAGP}\rangle = \frac{1}{N!} \left( \Gamma^\dagger \right)^N |\Downarrow\rangle,
\qquad
\Gamma^\dagger = \sum_p \eta_p\, S_p^+,
\end{equation}

where $S_p^+$ is the spin-raising operator at site $p$ and $|\Downarrow\rangle$ denotes the reference state with all spins down. This construction mirrors the fermionic AGP ansatz while replacing pair creation operators with spin excitations.

The resulting ansatz retains number and spin projection while maintaining mean-field computational cost. The wavefunction coefficients factorize as products of site amplitudes, reflecting the underlying symmetric-polynomial structure of AGP, and enabling efficient evaluation. The resulting ansatz captures nontrivial quantum phases in one- and two-dimensional spin systems and can be systematically improved by incorporating Jastrow-type correlators.

Closely related ideas appear in projected quasiparticle theory (PHFB), where symmetry-breaking Hartree–Fock–Bogoliubov states are projected variationally to restore particle number, spin, and other molecular symmetries.\cite{scuseria2011projected} From the geminal perspective this approach can be viewed as an extension of AGP with unrestricted quasiparticle orbitals. The variation-after-projection framework produces intrinsically multireference wavefunctions while retaining mean-field computational scaling, enabling efficient descriptions of strong correlation in molecular dissociation and other challenging regimes.

AGP-based approaches have also been extended to finite-temperature many-body theory. Harsha, Henderson, and Scuseria introduced a canonical-ensemble formulation based on thermofield dynamics in which the thermal density matrix is represented by a wavefunction in an enlarged Hilbert space.\cite{harsha2020wave} By projecting the grand-canonical thermal BCS state onto a fixed particle number, they showed that the canonical mean-field thermal state is formally equivalent to an AGP wavefunction with temperature-dependent geminal coefficients. Within this framework, the thermal state obeys an imaginary-time evolution equation analogous to the Schr\"odinger equation, enabling systematic inclusion of correlation effects beyond the mean-field approximation. Both projection-after-correlation schemes, such as projected configuration interaction, and correlation-after-projection approaches based on perturbation theory were explored. Benchmark calculations for systems including H$_2$ and small Hubbard clusters demonstrate that these methods recover a significant correlation missing from the thermal mean-field description while correctly reproducing differences between canonical and grand-canonical thermodynamics. These developments highlight that AGP provides a natural pairing reference not only for ground-state electronic structure but also for finite-temperature quantum many-body treatments.

Important computational advances have also improved the practical tractability of AGP-based methods. Khamoshi, Henderson, and Scuseria derived polynomial-cost algorithms to evaluate AGP norms and reduced density matrices (RDM) of arbitrary rank.\cite{khamoshi2019efficient} By expressing AGP RDM elements in terms of elementary symmetric polynomials and employing the numerically stable \texttt{sumESP} algorithm, individual RDM elements can be computed with quadratic cost, allowing calculations involving thousands of orbitals. Moreover, higher-rank AGP RDMs can be reconstructed from lower-rank occupation-number RDMs and geminal coefficients, greatly reducing the cost of correlated AGP-based approaches such as AGP-CI.

Beyond algorithmic developments, there have been efforts to understand the algebraic formulation of the wavefunctions in geminal basis. Recent work by \cite{sorensen2024transformation} reformulates electronic structure theory entirely in a geminal basis, where both the Hamiltonian and wavefunction are expressed in terms of pair creation and annihilation operators\cite{sorensen2024transformation}. Within this framework, the exact wavefunction is shown to satisfy a generalized Brillouin theorem, expressed as a vanishing expectation value of geminal replacement operators. This provides a rigorous variational condition for optimal geminal wavefunctions, analogous to the role of orbital rotations in Hartree–Fock theory. Importantly, the non-bosonic commutation relations of geminals introduce additional coupling terms, implying that common ansätze such as AGP and APG are not invariant under general geminal rotations and therefore cannot satisfy the exact stationary conditions. In contrast, the full geminal product (FGP) ansatz,

\begin{equation}
    |\Psi_{\text{FGP}} \rangle = \sum_{\nu_1 \leq \nu_2 \leq ... \nu_{N/2}} C_{\nu_1 \nu_2 \hdots \nu_{N/2}} \psi_{\nu_1}^\dagger \psi_{\nu_2}^\dagger
    \hdots\psi_{\nu_{N/2}}^\dagger | \ \rangle = \sum_p C_p \Psi_p
\end{equation}
spans all possible products of geminals and is therefore complete. As a result, it satisfies the exact stationary conditions and is formally equivalent to full configuration interaction expressed in a geminal basis.

These developments collectively illustrate how AGP serves not only as a practical pairing ansatz but also as a conceptual bridge connecting geminal wavefunctions, reduced-density-matrix formulations, and broader many-body pairing theories.

\subsection{1-reference Orbital Approximation}

While pair-product wavefunctions such as APIG, APSG, and AGP provide compact descriptions of electron pairing, their direct evaluation or optimization can remain computationally demanding. A practical strategy to alleviate this difficulty is to retain the APIG pairing structure while introducing a reference determinant, which enables a coupled-cluster–like parameterizations of the geminal coefficients.

Limacher \textit{et al.}\cite{limacher2013new} introduced the \acrfull{ap1rog}, providing the first computationally practical realization of interacting non-orthogonal geminal wavefunctions derived from the APIG framework. In this approach, each geminal is associated with a unique occupied spin-orbital pair (highlighted in eq.(\ref{eq:ref-orb})) taken from a reference Slater determinant. This introduces the occupied-virtual orbital distinction within the geminal formalism. Electron correlation is then incorporated through pair excitations from the reference pairs into the virtual orbitals space.

This restriction significantly simplifies the structure of the geminal coefficient matrix and allows the projected Schr\"odinger equation to be solved with polynomial computational cost. As a consequence, the method exhibits near mean-field computational scaling of $\mathcal{O}(P^2(K-P)^2)$, where $P=N/2$ is the number of electron pairs and $K$ the number of spatial orbitals. 

\begin{equation}\label{eq:ref-orb}
    \hat{G}_i^\dagger = \underset{\substack{\downarrow \\ \text{unique occupied} \\ \text{spin-orbital pair}}}{a_i^\dagger a_{\bar{i}}^\dagger} + \sum_a^{\text{virt}} C_i^a a_a^\dagger a_{\bar{a}}^\dagger
\end{equation}

\begin{equation} \label{eq:pccd==ap1rog}
    \begin{split}
            | \Psi_{\text{pCCD}} \rangle & = \prod_{i=1}^{P=N/2} \left( a_i^\dagger a_{\bar{i}}^\dagger + \sum_a^{\text{virt}} C_i^a a_a^\dagger a_{\bar{a}}^\dagger \right) | \ \rangle = \prod_{i=1}^{P=N/2} \left( 1 + \sum_a^{\text{virt}} C_i^a a_a^\dagger a_{\bar{a}}^\dagger a_{\bar{i}} a_i \right) a_i^\dagger a_{\bar{i}}^\dagger | \ \rangle \\
            & = \exp \left( \sum_i^{\text{occ}} \sum_a^{\text{virt}} C_i^a a_a^\dagger a_{\bar{a}}^\dagger a_{\bar{i}} a_i\right) |\Phi_0\rangle = \exp (\hat{T_p}) | \Phi_0 \rangle
    \end{split}
\end{equation}

Despite its compact form, AP1roG reproduces the results of the seniority-zero configuration interaction (DOCI) wavefunction with high accuracy while remaining computationally tractable. The method is formally equivalent to the pair coupled-cluster doubles (pCCD) model (eq.(\ref{eq:pccd==ap1rog})), revealing a deep connection between geminal wavefunctions and coupled-cluster theory. Benchmark studies have demonstrated that AP1roG captures the dominant static correlation in strongly correlated systems such as hydrogen chains while retaining the near mean-field computational cost. 

A detailed discussion of the algebraic simplifications and graphical representations of the coefficient structures for various pair-product wavefunctions, including APIG, Richardson–Gaudin states, AGP, pCCD, APSG, and GVB-PP—can be found in the review by Tecmer and Boguslawski.\cite{tecmer2022geminal}

A central challenge in AP1roG and related methods is the need for orbital optimization to identify the appropriate electron-pairing structure. When the orbitals are fully optimized, AP1roG becomes size-consistent, as demonstrated in the benchmark studies of Limacher \textit{et al.}

Subsequent work by Boguslawski \textit{et al.}\cite{boguslawski2014efficient} developed an orbital-optimized formulation of AP1roG (OO-AP1roG), in which the one-electron basis is variationally optimized together with the geminal coefficients. Orbital rotations spanning the occupied–occupied, occupied–virtual, and virtual–virtual subspaces are optimized using a Newton–Raphson procedure, analogous to orbital-optimized coupled-cluster approaches. While the geminal amplitude equations retain mean-field-like scaling, the required transformation of two-electron integrals increases the overall computational cost to $\mathcal{O}(K^5)$. Benchmark studies on strongly correlated systems such as the one-dimensional Hubbard model and the symmetric dissociation of hydrogen chains demonstrate that OO-AP1roG accurately captures a large fraction of the correlation energy and yields potential energy curves in close agreement with density-matrix renormalization group (DMRG) reference results. These developments establish orbital optimization as a crucial component for obtaining reliable and size-consistent AP1roG descriptions of strongly correlated systems.

Further insight into the role of orbital optimization in closed-shell geminal methods was provided by Limacher \textit{et al.}\cite{limacher2014influence}. By analyzing doubly-occupied configuration interaction (DOCI), the most general seniority-zero wavefunction, the authors demonstrated that the energy of closed-shell wavefunctions depends strongly on the chosen orbital basis and that the orbital optimization landscape may contain multiple local minima corresponding to different electron-pairing patterns. Importantly, they showed that DOCI correctly describes bond dissociation already with restricted orbitals, in contrast to Hartree–Fock where symmetry breaking is often required. The study further confirmed that the AP1roG wavefunction closely reproduces DOCI energies, typically differing by less than a millihartree for the systems considered, thereby supporting AP1roG as an efficient approximation to the computationally expensive DOCI method.

The conceptual foundation for seniority-based wavefunction models was established by Bytautas \textit{et al.}, who proposed partitioning the full configuration interaction (FCI) Hilbert space according to the seniority number $\Omega$, defined as the number of unpaired electrons in a Slater determinant.\cite{bytautas2011seniority} Within this hierarchy, the $\Omega=0$ sector contains determinants with all electrons paired and corresponds to the doubly-occupied configuration interaction (DOCI) wavefunction. Their analysis demonstrated that the low-seniority sectors capture the dominant static correlation in strongly correlated systems, while higher-seniority contributions primarily account for dynamic correlation. This observation provides a theoretical rationale for the success of geminal-based approaches such as AP1roG, which efficiently parametrize the seniority-zero sector of the Hilbert space.

Building on these ideas, Boguslawski \textit{et al.}\cite{boguslawski2014projected} introduced a projected seniority-two orbital optimization scheme for AP1roG. The approach exploits the observation that orbital-optimized seniority-zero and seniority-zero-plus-two configuration interaction expansions yield similar energies when appropriate orbitals are used. By enforcing the decoupling condition between the seniority-zero and seniority-two sectors, the method provides a computationally efficient orbital optimization strategy that preserves spatial symmetry while maintaining the accuracy of the AP1roG description.


Recent developments have also explored geminal–coupled-cluster hybrids. Ahmadkhani \textit{et al.} demonstrated that coupled-cluster doubles of linear-response pairs with singles (LR-pCCD + S), using canonical Hartree–Fock orbitals, achieves favorable scaling relative to EOM-CCSD while maintaining accuracy for excited states in extended $\pi$-systems such as transpolyenes.\cite{ahmadkhani2024linear} These developments illustrate the adaptability of geminal concepts for describing both ground- and excited-state correlation.

The practical applicability of geminal-based wavefunctions to large $\pi$-conjugated systems has been recently highlighted in the context of organic electronic materials. Tecmer and co-workers demonstrated that orbital-optimized pCCD (AP1roG) combined with equation-of-motion (EOM) formalisms provides a scalable and systematically improvable alternative to density functional approximations for modeling organic photovoltaics (OPVs) and dye-sensitized solar cells\cite{tecmer2023geminal}. In contrast to conventional DFAs, which often misrepresent delocalization, torsional barriers, and charge-transfer states in extended $\pi$-systems, geminal-based models offer a compact multireference description with polynomial scaling. A central challenge addressed in this work concerns the extraction of physically meaningful orbital energies from natural-orbital-based geminal theories. By employing IP/EA-EOM-pCCD and extended Koopmans’ theorem formulations, charge gaps ($\Delta$c = IP - EA) and related electronic properties can be computed from reduced density matrices. Benchmarks on representative acceptor and donor–$\pi$–acceptor dyes showed that inclusion of higher excitation sectors (2h1p/2p1h) substantially improves predicted HOMO–LUMO gaps compared to simpler 1h/1p treatments and even competes favorably with DLPNO-CCSD in some cases. Beyond frontier orbital energies, the authors emphasized that the localized nature of pCCD-optimized orbitals enables an unambiguous assignment of donor and acceptor domains and facilitates automated analysis of charge-transfer character in excited states. This interpretability, combined with efficient evaluation of one- and two-particle reduced density matrices, further allows the computation of quantum entanglement measures, non-covalent interaction energies, and embedding schemes tailored for large molecular aggregates. Altogether, these developments position geminal-based methods as promising candidates for scalable, property-oriented modeling of large organic electronic building blocks.

Limacher \textit{et al.} subsequently proposed hybrid models such as the antisymmetrized product of rank-two reference-orbital geminals (APr2roG)\cite{limacher2013new}. In this approach, the structured rank-two parametrization of APr2G is retained for the reference orbital pairs, while the remaining coefficients associated with virtual orbital pairs are treated more flexibly, analogous to the parametrization used in AP1roG. This hybrid construction provides a systematic route for interpolating between highly structured rank-two geminal forms and more flexible interacting-pair descriptions, thereby improving the practical applicability of the approach.

Building upon the one-reference-orbital idea, Gaikwad \textit{et al.} recently introduced a family of coupled-cluster–inspired geminal wavefunctions that extend the 1-reference-orbital formalism to more general geminal ans\"atze such as APG and APsetG.\cite{gaikwad2024coupled} These models retain the compact quasiparticle structure of geminal wavefunctions while exploiting the exponential parametrization characteristic of coupled-cluster theory. In this framework, several new wavefunction hierarchies emerge that behave as approximations to CCD but remain robust in strongly correlated regimes where conventional single-reference methods fail. In efforts to capture the missing dynamical correlation in the 1-reference orbital models and to alleviate the difficulty of orbital optimization, the authors introduced ``faux singles'' operators highlighted in the following equation:

\begin{equation}\label{eq:g_ap1rogsd}
    G^\dagger_{i(\text{AP1roGSD})} = a_i^\dagger a_{\bar{i}}^\dagger + \sum_a c_{i\bar{i}}^{a\bar{a}} a_a^\dagger a^\dagger_{\bar{a}} +   \underbrace{\sum_a c_{i\bar{i}}^{a\bar{i}} a_a^\dagger a^\dagger_{\bar{i}} +   \sum_a c_{i\bar{i}}^{i\bar{a}} a_{\bar{a}}^\dagger a^\dagger_{i}}_{\text{creation operators contributing faux singles}},
\end{equation}
which are geminal-preserving double excitations that mimic the role of $T_1$ amplitudes in coupled-cluster theory. By substituting the above operator $G^\dagger_{i(\text{AP1roGSD})}$ (eq. \ref{eq:g_ap1rogsd}) in the general antisymmetrized product form in eq.(\ref{eq:general_geminal}), and following the simplification of eq.(\ref{eq:pccd==ap1rog}), we obtain the geminal wavefunction, eq.(\ref{eq:ap1rogsd}),  with doubles and singles configurations. These operators effectively induce orbital relaxation while preserving the underlying two-electron geminal picture, enabling more robust and black-box implementations of geminal wavefunctions.

\begin{equation}\label{eq:ap1rogsd}
    | \Psi_{\text{AP1roGSD}} \rangle = \exp \left( \sum_{i,a} c_{i\bar{i}}^{a\bar{a}} a_a^\dagger a^\dagger_{\bar{a}} a_{\bar{i}} a_i +  \sum_{i,a} c_{i\bar{i}}^{a\bar{i}} a_a^\dagger a^\dagger_{\bar{i}} a_{\bar{i}} a_i +   \sum_{i,a} c_{i\bar{i}}^{i\bar{a}} a_{\bar{a}}^\dagger a^\dagger_{i} a_{\bar{i}} a_i\right) | \Phi_{0} \rangle
\end{equation}

The work further explored different strategies for introducing single-like excitations through controlled breaking of seniority constraints (sen-o, sen-v, sen-ov, and sen-free variants). Numerical tests on strongly correlated model systems, including hydrogen clusters and the perpendicular insertion of Be into H$_2$, demonstrate that maintaining a strict geminal structure while incorporating these pseudo-single excitations provides a promising route toward practical orbital-optimization-free geminal methods.


Together, these developments demonstrate that imposing structured constraints on the geminal coefficients such as reference-orbital restrictions or low-rank parameterizations, provides a powerful route to transforming formally intractable interacting geminal wavefunctions into computationally practical models with coupled-cluster–like scaling.

\subsection{Richardson–Gaudin States}

Richardson–Gaudin (RG) states provide a structured and computationally tractable realization of interacting geminal wavefunctions. In the most general antisymmetrized product of interacting geminals (APIG), the expansion coefficients in a Slater determinant basis are matrix permanents, rendering the wavefunction computationally intractable. A key simplification arises by imposing a Cauchy structure on the geminal coefficients, which defines the RG wavefunction and allows permanents to be evaluated through ratios of determinants using Borchardt’s theorem.

The Richardson–Gaudin (RG) state $|\{u\}\rangle$ is therefore written as an antisymmetrized product of collective geminal creation operators $S^+(u_\alpha)$,
\begin{equation}
|\{u\}\rangle = \prod_{\alpha=1}^P S^+(u_\alpha)|\theta\rangle,
\end{equation}

where each operator is a linear combination of local pair-creation operators $S_i^\dagger = a_{i\uparrow}^\dagger a_{i\downarrow}^\dagger$,

\begin{equation}
S^+(u_\alpha)=\sum_i \frac{S_i^\dagger}{u_\alpha-\varepsilon_i}.
\end{equation}

Here $\varepsilon_i$ denote orbital energies and $u_\alpha$ are complex parameters known as rapidities or pair energies.

When the rapidities satisfy the nonlinear Richardson equations, the resulting RG state becomes an exact eigenstate of the reduced \acrfull{bcs} pairing Hamiltonian. Even when treated variationally for general Hamiltonians, the Cauchy structure of the coefficients enables determinant expressions for overlaps and reduced density matrices, making RG states a computationally feasible approximation to the otherwise intractable APIG wavefunction.

RG states originate from the algebraic Bethe Ansatz solution of integrable pairing Hamiltonians. When the rapidities $\{u_\alpha\}$ satisfy the nonlinear Richardson equations, the resulting wavefunction is an exact eigenstate of the reduced \acrshort{bcs} Hamiltonian. Such solutions are referred to as \emph{on-shell} Richardson–Gaudin states, where the rapidities are constrained so that residual pair-scattering terms cancel. In contrast, \emph{off-shell} RG states relax this constraint and treat the rapidities as independent variational parameters. This variational extension preserves the favorable algebraic structure of the RG ansatz while enlarging the accessible wavefunction manifold beyond the strictly integrable limit.

Fecteau \textit{et al.}\cite{fecteau2021richardson} reformulated RG geminal wavefunctions in the conventional Slater determinant language of quantum chemistry, clarifying their relationship to pair coupled-cluster and AP1roG/pCCD approaches. They derived compact closed-form expressions for RG cumulants in the Hartree–Fock determinant basis and demonstrated how cluster amplitudes may be reconstructed, yielding a structured hierarchy of determinant-based approximations to APIG (denoted by the abbreviation pCCdr, for pair-coupled cluster determinant ratio). In this framework, RG states emerge as mathematically controlled approximations to interacting geminal theories, preserving the integrable pairing structure while enabling systematic improvements toward full APIG.

Beyond formal connections to coupled-cluster theory, the flexibility of RG states as variational ans\"atze for nonintegrable Hamiltonians has been explored. Claeys \textit{et al.}\cite{claeys2017variational} proposed optimizing \emph{on-shell} RG eigenstates with respect to integrability-breaking perturbations. Because Bethe-ansatz RG states admit determinant expressions for overlaps and expectation values, the resulting energy functional can be minimized at polynomial cost. For Hamiltonians close to an integrable pairing limit, the variationally optimized RG ground state substantially improves upon first-order perturbation theory and captures dominant pairing correlations. When integrability-breaking interactions induce avoided level crossings, accurate descriptions may require variational optimization of RG excited states rather than the integrable ground state, delineating the regime of applicability of integrability-based wavefunctions.

Recent work has further explored the use of RG states as variational ans\"atze for strongly correlated electronic systems. Johnson \textit{et al.}\cite{johnson2020richardson} proposed a Richardson–Gaudin mean-field framework in which eigenvectors of the reduced BCS Hamiltonian are employed as trial wavefunctions for the Coulomb Hamiltonian. In this formulation, the RG state acts as a mean-field description of interacting electron pairs, analogous to how Hartree–Fock describes weakly correlated electrons. Benchmark calculations for atoms and molecular dissociation show that a single optimized RG state can recover a large fraction of the correlation captured by seniority-zero methods such as DOCI while retaining polynomial computational scaling.

A complementary and conceptually unifying perspective has been provided by Johnson \textit{et al.}\cite{johnson2025connections}, who established explicit connections between RG states, perfect pairing (PP), and pair coupled-cluster doubles (pCCD). Reformulating the reduced BCS Hamiltonian in a bonding/antibonding representation, they showed that PP arises as the independent-pair limit of RG theory. Each pair is characterized by a gap parameter that quantifies the competition between orbital energy splitting and pair-transfer interaction, providing a transparent diagnostic of weak versus strong correlation. Second-order Epstein–Nesbet perturbation theory on top of orbital-optimized PP yields energies nearly identical to pCCD and seniority-zero DOCI for prototypical hydrogen chains, and in the single-pair limit PP and pCCD become formally equivalent. This analysis positions PP as a mean-field reduction of RG states and clarifies how pCCD systematically incorporates residual inter-pair correlation beyond the independent-pair picture.

The extension of RG geminals beyond strict integrability has recently been formalized through a bivariational framework.\cite{johnson2022bivariational} In this formulation, RG-type wavefunctions are expressed in dual particle and pair-hole representations,
\begin{equation}
|\Psi(\{v\})\rangle
=
\prod_{\alpha=1}^{P} S^+(v_\alpha)|\theta\rangle
=
\prod_{\beta=1}^{K-P} S(\tilde u_\beta)|\tilde{\theta}\rangle ,
\end{equation}
where $S^+(v_\alpha)$ creates a collective pair acting on pair vacuum state $| \theta \rangle$ and $S(\tilde u_\beta)$ annihilates a pair (hole picture) from the fully occupied reference $| \tilde{\theta} \rangle$. The two representations are treated as independent variational objects and optimized through a bivariational principle. When the particle and pair-hole representations coincide, the energy becomes strictly variational. Importantly, even for these off-shell, nonorthogonal geminal constructions, reduced density matrices remain expressible as ratios of determinants via Borchardt’s theorem, preserving polynomial scaling. The framework encompasses hyperbolic (XXZ-type) generalizations and explicitly nonorthogonal orbitals, and recovers APSG/GVB-PP, AGP, and on-shell RG states as special cases.

Johnson and DePrince investigated the performance of variational RG states for strongly correlated hydrogen model systems, focusing on the Paldus H$_4$ isomers (square, rectangular, linear, and bond-swapping geometries) and the Stair–Evangelista H$_{10}$ isomers (1D chain, 1D ring, 2D sheet, and 3D pyramid) in a minimal STO-6G basis\cite{johnson2023single}. These systems provide controlled multireference testbeds that mimic finite-size Hubbard-like models with varying connectivity and dimensionality. Across all cases, a single variationally optimized RG state accurately reproduces the seniority-zero sector as described by orbital-optimized DOCI (OO-DOCI), typically within a few millihartree over full potential energy curves. Systematic improvement is achieved by performing configuration interaction in the RG basis (RGCIS and RGCISD), which reduces residual errors by several orders of magnitude and essentially recovers OO-DOCI. The study further showed that different bonding regimes correspond to distinct optimal RG bitstrings (e.g., Néel-type versus block-partitioned structures), reflecting reorganizations of collective pair structure as geometry changes. Simple occupation-number-based weak-correlation corrections were also tested, providing modest geometry-dependent improvements. Overall, this work establishes RG states as compact, variationally controllable, and systematically improvable single-reference ans\"atze capable of describing strong pairing correlations across one-, two-, and three-dimensional hydrogen networks.

Taken together, these developments position RG geminals at the intersection of integrable models, coupled-cluster theory, and variational many-body methods. They furnish both a formally controlled reduction of APIG and a physically motivated pathway for describing strongly correlated systems dominated by pairing structure.


\subsection{Generalized Valence Bond Method}
%
%

The \acrfull{gvb} method was developed by William A. Goddard III to overcome limitations of the \acrshort{hf} wavefunction in describing bond dissociation and other strongly correlated situations. A detailed historical overview of the development and impact of GVB theory is provided in a recent book chapter by Dunning and Hay.\cite{Dunning2021} In contrast to the single-determinant HF description, GVB explicitly represents electron pairing through optimized two-electron functions, allowing non-dynamical correlation to be incorporated at the mean-field level.

In GVB theory the wavefunction is expressed as an antisymmetrized product of pair functions (geminals),

\begin{equation}
\Psi_{\text{GVB}} = \mathcal{A} \left[ \prod_{p=1}^{P} \phi_i(\textbf{r}_i)\right] \Theta_{S,M}
\end{equation}

where $\mathcal{A}$ is the antisymmetrizer, $\{\phi_i\}$ are spatial orbitals, and $\Theta_{S,M}$ is a general spin eigenfunction describing the coupling of electron spins to total spin $S$ and projection $M$. In this most general form, the orbitals may be non-orthogonal and the spin coupling among electrons is treated variationally.

The complexity of the full GVB equations motivated simplified forms of the theory. The most widely used approximation is its perfect-pairing model - \acrfull{gvb-pp}, in which electrons are assumed to form independent singlet-coupled pairs. In this approximation the general spin function $\Theta_{S,M}$ is restricted to the perfect-pairing form

\begin{equation}
\Theta_{\text{PP}} =
\prod_{p=1}^{P}
\frac{1}{\sqrt{2}}
\left(
\alpha_p \beta_p - \beta_p \alpha_p
\right),
\end{equation}

which describes a product of singlet spin functions for each electron pair. The corresponding wavefunction can then be written as an antisymmetrized product of two-electron geminals,

\begin{equation}
\Psi_{\text{GVB-PP}} =
\mathcal{A} \prod_{p=1}^{P} g_p(1,2),
\end{equation}

where each geminal describes a singlet electron pair,

\begin{equation}
g_p(1,2) =
\frac{1}{\sqrt{2}}
\left[
\phi_p(1)\psi_p(2) - \psi_p(1)\phi_p(2)
\right]
\chi_{\text{singlet}} .
\end{equation}

A further simplification is obtained by imposing the \textit{strong orthogonality} (SO) condition between orbitals belonging to different pairs,

\begin{equation}
\langle \phi_p | \phi_q \rangle =
\langle \psi_p | \psi_q \rangle =
\langle \phi_p | \psi_q \rangle = 0
\qquad (p \neq q),
\end{equation}

which yields the GVB-PP/SO wavefunction. This constraint eliminates interpair overlap and greatly simplifies the working equations while retaining the chemically intuitive description of localized electron pairs.

Within this hierarchy of approximations, the Hartree–Fock wavefunction can be viewed as a limiting case of GVB-PP/SO in which the two orbitals within each pair are constrained to be identical ($\phi_p=\psi_p$), reducing each geminal to a doubly occupied orbital. Thus the sequence

\[
\Psi_{\text{GVB}}
\;\supset\;
\Psi_{\text{GVB-PP}}
\;\supset\;
\Psi_{\text{GVB-PP/SO}}
\;\supset\;
\Psi_{\text{HF}}
\]

represents a progressive restriction of the geminal wavefunction toward the familiar single-determinant description.


GVB has proven particularly useful as an interpretive framework for understanding unconventional bonding patterns. Xu and Dunning employed fully variational GVB calculations to analyze the controversial bond order of the diatomic carbon molecule, C$_2$.\cite{xu2014insights} While C$_2$ has been variously described as having double, triple, or even quadruple bonding, their spin-coupled analysis showed that the dominant configuration corresponds to a single conventional $\sigma$ bond with the remaining valence electrons on each carbon atom anti-ferromagnetically coupled in a quasi-atomic fashion. Although perfect-pairing configurations contribute significantly, strong Pauli repulsion between overlapping $\sigma$ pairs suppresses the formation of four distinct shared pairs. This study illustrates the interpretive power of GVB in disentangling non-dynamical correlation effects that are often obscured in orbital-based descriptions.

The work by Voorhis and Head-Gordon demonstrates that GVB pair-based wavefunctions can be expressed in coupled-cluster form. In particular, the perfect-pairing (PP) wavefunction can be written as 
\begin{equation}\label{eq:gvb-ppcc}
    | \Psi_{PP}\rangle = e^{\hat{T}_{PP}} | \Phi \rangle,
\end{equation}
where the cluster operator excites an electron pair from an occupied orbital $i$ to its corresponding virtual orbital $i^*$,
\begin{equation}
    \hat{T}_{PP} = \sum_i t_{i\bar{i}}^{i^* \bar{i}^*} a^\dagger_{i^*} a^\dagger_{\bar{i}^*} a_{\bar{i}} a_i.
\end{equation}

However, when the GVB restricted configuration interaction (GVB-RCI) expansion is represented in a conventional CC exponential form, additional terms appear that violate an effective pairwise exclusion principle present in the GVB formulation. Removing these terms leads to the modified GVB-RCC approach, which yields improved dissociation behavior for strongly correlated systems.\cite{van2001connections}

Several extensions have been proposed to improve the description of dynamical correlation within the perfect-pairing framework. In particular, Beran, Head-Gordon, and co-workers developed an unrestricted perfect-pairing (UPP) formulation that generalizes the coupled-cluster representation of perfect pairing to open-shell systems.\cite{beran2005unrestricted} In UPP, the same PP coupled-cluster ansatz of Eq.~\eqref{eq:gvb-ppcc} is employed, but the spatial $\alpha$ and $\beta$ orbitals are optimized independently. The occupied space is partitioned into core, paired active orbitals, and singly occupied orbitals, where only the paired electrons are correlated through the PP cluster operator while the unpaired electrons are treated in a manner analogous to unrestricted Hartree–Fock. This formulation preserves the correct dissociation behavior and improves molecular structures and radical descriptions relative to HF, while retaining the favorable linear scaling characteristic of pair-based theories.

A perturbative correction to the perfect-pairing reference was later introduced by Beran, Head-Gordon, and Gwaltney through the PP(2) method.\cite{beran2006second} Using similarity-transformed perturbation theory applied to the coupled-cluster representation of PP, the authors derived an MP2-like correction that restores dynamical correlation while preserving the strong bond-breaking capabilities of the PP reference. PP(2) achieves accuracy comparable to MP2 for many thermochemical properties while providing improved potential energy curves in systems exhibiting significant static correlation, such as stretched N$_2$ and diradicaloid species.

An alternative strategy for restoring dynamical correlation in pair-based wavefunctions employs response-theoretic corrections derived from reduced density matrices. Chatterjee \textit{et al.} introduced an extended random phase approximation (ERPA) correction to the GVB-PP reference, in which the missing correlation energy is expressed through fluctuations of the one- and two-electron reduced density matrices obtained from ERPA response equations.\cite{chatterjee2016minimalistic} The resulting ERPA-GVB framework introduces both intra- and intergeminal correlation contributions while retaining the compact perfect-pairing reference. 

Subsequent developments generalized this approach through an embedding formulation termed EERPA-GVB.\cite{pastorczak2019generalized} In this scheme, orbitals are partitioned into fragments corresponding to individual geminals or groups of overlapping pairs, and the correlation energy is decomposed into intra- and inter-fragment contributions. Truncation of the ERPA expansion to one- and two-fragment terms yields a computational cost that scales approximately quadratically with the number of electron pairs while recovering both short-range dynamical correlation and long-range dispersion interactions. Applications to reaction barriers, weakly interacting complexes, and conjugated systems demonstrate that the embedding formulation significantly improves upon the GVB-PP reference while preserving the chemically transparent electron-pair description.

More recently, Nakatani and Sato proposed a generalized pairing (GP) geminal framework that relaxes the singlet-pairing restriction by allowing mixing of singlet and triplet components within each geminal.\cite{nakatani2023geminal} 
Within the pair-product formalism introduced in Eq.~(\ref{eq:general_geminal}), the geminal creation operator is generalized to include all spin combinations,
\begin{equation}
\hat{G}_p^\dagger =
\sum_{q<r} \sum_{\sigma_1 \sigma_2}
C_{qr}^{p,\sigma_1\sigma_2}
\, a^\dagger_{q\sigma_1} a^\dagger_{r\sigma_2},
\end{equation}
where the coefficients $C_{qr}^{p,\sigma_1\sigma_2}$ allow singlet and triplet components to mix within each geminal. 

Formulated as an antisymmetrized product of strongly orthogonal geminals, the GP ansatz may be viewed as a spin-generalized analogue of GVB-PP and APSG, analogous to the extension from restricted to generalized Hartree–Fock in single-determinant theories. Variational optimization of both the orbitals and geminal coefficients yields broken-symmetry solutions that capture intra-geminal correlation more effectively than conventional perfect-pairing models.

Further systematic improvements have been explored through block-correlated correlation methods based on GVB references. In these approaches, each geminal is treated as a correlated “block” and inter-block excitations are introduced to recover dynamical correlation beyond the perfect-pairing picture. Early developments include block-correlated second-order perturbation theory based on GVB references (GVB-BCPT2), in which intra-block correlation is treated exactly within each geminal while inter-block interactions are incorporated perturbatively through simultaneous block excitations.\cite{xu2013block} This framework provides a size-consistent MP2-like correction applicable to systems with large active spaces that are difficult for traditional CASSCF-based perturbation methods.

Building on the same block-partitioning philosophy, Han and co-workers developed block-correlated coupled-cluster formulations based on GVB references (GVB-BCCC), which systematically incorporate correlations among multiple electron pairs.\cite{ren2024block} High-order variants such as GVB-BCCC5 include correlations among up to five pairs and have been successfully applied to strongly correlated systems including hydrogen lattices, phosphorus clusters, and transition-metal oxides. Comparisons with \acrfull{dmrg} results demonstrate that these block-correlated approaches can achieve high accuracy while maintaining a compact pair-based description of the wavefunction.

Time-dependent extensions of strongly orthogonal geminal models have also been developed to access excited-state properties. Hapka, Pernal, and Jensen implemented an efficient time-dependent linear-response formalism for GVB-PP and APSG wavefunctions, enabling the calculation of singlet excitation energies within a geminal framework.\cite{hapka2022efficient} The approach combines a second-order optimization algorithm for geminal parameters with direct iterative solution of the response equations. Benchmark studies on small and medium-sized molecules show that TD-GVB improves slightly over time-dependent Hartree–Fock but remains less accurate than linear-response CASSCF, particularly for states with significant double-excitation character.

\begin{figure}[h]
    \centering
    \includegraphics[width=0.99\linewidth]{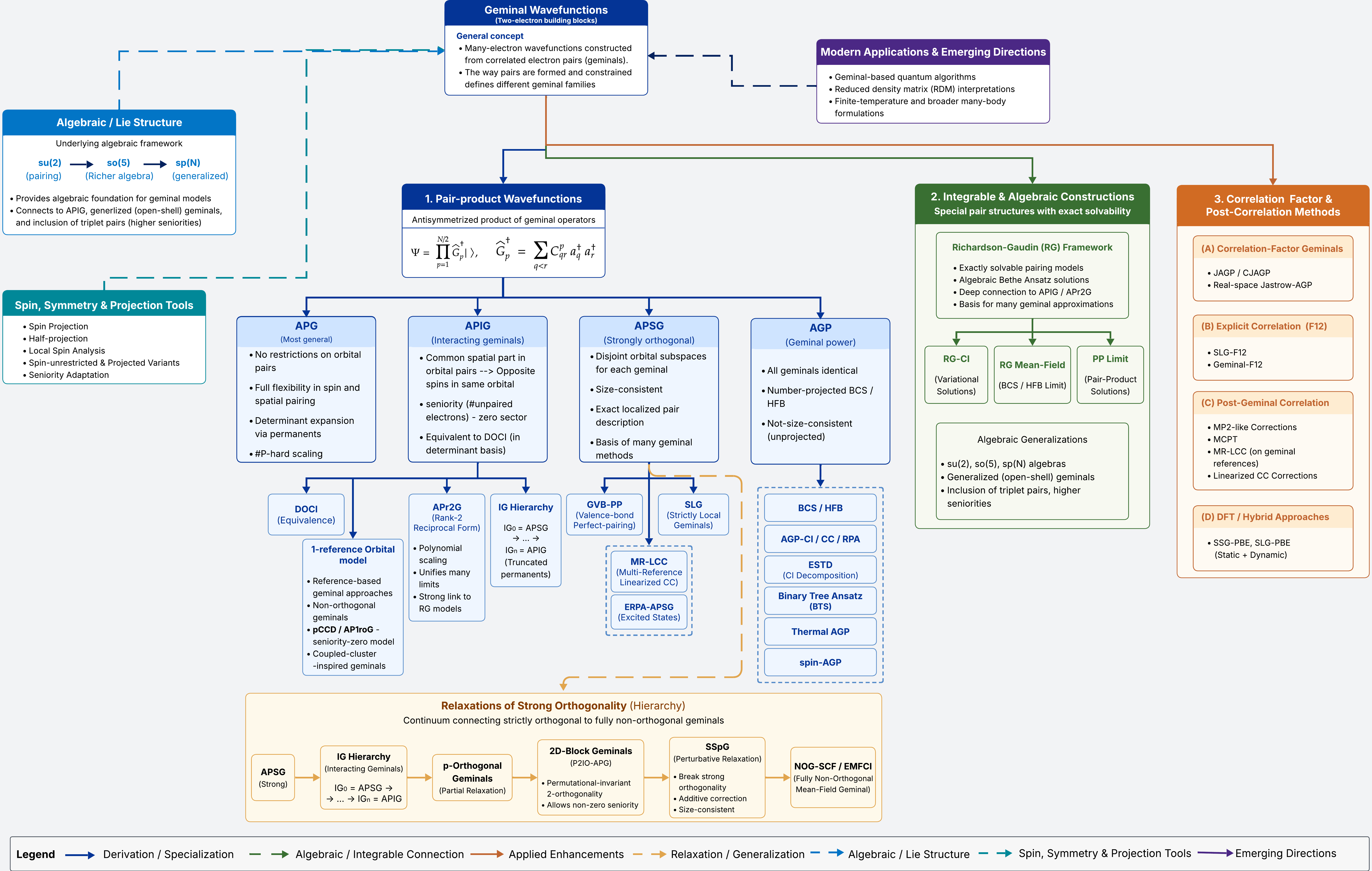}
    \caption{Schematic overview of geminal wavefunction theory highlighting the main conceptual relationships and methodological developments discussed in this mini-review. The diagram organizes representative classes of pair-product wavefunctions, algebraic and integrable constructions, correlation treatments, and related extensions into a unified framework. The connections shown are intended to provide a conceptual roadmap rather than an exhaustive classification; many additional developments and variations exist beyond those included here.}
    \label{fig:Conceptual Diagram of various branches of studies of geminal wavefunctions.}
\end{figure}

\section{Correlation-Factor Geminals}

Beyond variational geminal wavefunctions such as AGP, geminal concepts also play a central role in explicitly correlated (R12/F12) and transcorrelated (TC) methods, which aim to incorporate short-range electron correlation directly into the theoretical framework and thereby accelerate basis-set convergence. In explicitly correlated approaches, the wavefunction is augmented by terms that depend explicitly on the inter-electronic distance $r_{12}$, allowing it to satisfy the electron–electron cusp condition and recover dynamic correlation more efficiently than conventional orbital expansions. An alternative strategy is provided by transcorrelated methods, in which correlation is incorporated through a similarity transformation of the Hamiltonian rather than by explicitly modifying the wavefunction ansatz.

\subsection{Explicitly Correlated Geminals}


The slow convergence of conventional electronic structure methods with respect to one-particle basis sets originates from their inability to describe the electron–electron cusp. Explicitly correlated approaches address this limitation by introducing geminal functions that depend directly on the interelectronic distance $r_{12}$, thereby capturing short-range correlation effects efficiently.

Early developments in explicitly correlated methods also explored the use of Gaussian-type geminals (GTGs) to incorporate interelectronic distance dependence into correlated wavefunctions. Polly et al. introduced GTGs within a local MP2 framework, where pair functions are augmented by Gaussian correlation factors of the form $e^{-\gamma r^2_{12}}$\cite{polly2006application}. This approach significantly improves basis-set convergence, recovering over 90\% of the correlation energy with double-zeta basis sets. Although GTGs do not satisfy the electron–electron cusp condition exactly, they provide an efficient description of medium-range correlation effects. The main limitation of the method lies in the computational cost associated with three-electron integrals, which motivated later developments such as RI-based R12 and F12 methods.

A major step in the direction of explicitly correlated methods was taken by Ten-no, who introduced the use of Slater-type geminals (STGs), of the form $f(r_{12}) = \frac{e^{-\beta r_{12}}}{\beta}$, where $\beta$ is \textit{inverse length scale}, in explicitly correlated theory\cite{ten2004initiation}. Unlike the commonly used linear $r_{12}$ or Gaussian-type geminals, STGs provide a more physically appropriate description of the electron–electron cusp while remaining computationally tractable through resolution-of-the-identity (RI) techniques. This work laid the foundation for modern F12 methods by demonstrating improved numerical performance and faster convergence of correlation energies.

Building on this framework, F12 methods have become a standard tool for accelerating basis-set convergence in correlated calculations. In these approaches, the wavefunction is augmented with geminal terms that explicitly model the coalescence region, leading to near basis-set limit accuracy even with moderate basis sets\cite{tew2018relaxing}. Tew and Kats highlighted an important limitation of commonly used fixed-amplitude F12 schemes and introduced a Lagrangian-based correction that relaxes these constraints. Their approach improves orbital-optimized and core–valence correlation energies while retaining the low computational cost characteristic of F12 methods.

Further refinement of explicitly correlated models has focused on the choice and parametrization of the geminal function itself. Powell et al. revisited the standard Slater-type correlation factor and showed that geminal length-scale parameters optimized at the MP2-F12 level are suboptimal for higher-level methods such as CCSD-F12\cite{powell2025slimmer}. By introducing reoptimized (“slimmer”) geminals, they demonstrated significant reductions in basis-set incompleteness errors, particularly for absolute correlation energies and for basis sets specifically designed for F12 calculations. This work emphasizes that even within the single-geminal framework, careful parameter selection is crucial for achieving high accuracy.

In parallel, alternative formulations have explored how explicitly correlated geminals can be incorporated more directly into many-body formalisms. Bayne et al. developed a geminal-projected configuration interaction (GP-CI) framework in which a two-body R12 geminal operator is used to analytically identify and eliminate noncontributing configurations\cite{bayne2016construction}. By constructing geminal-projected particle–hole excitation operators through diagrammatic factorization, the method achieves substantial reductions in CI space while maintaining high accuracy. This approach highlights a different role of geminals—not only improving basis convergence, but also guiding compact wavefunction representations.

More recently, explicitly correlated techniques have been extended to geminal-based multireference frameworks, bridging static and dynamic correlation treatments. Mihálka and Noga introduced an explicitly correlated extension of geminal-based perturbation theory, in which a reference wavefunction constructed from strongly orthogonal geminals (SLG) captures static correlation, while dynamic correlation is recovered via a second-order perturbative treatment augmented with an F12 correction\cite{mihalka2025explicitly}. Notably, although individual geminals satisfy the electron–electron cusp condition, their antisymmetrized product does not, leading to residual basis-set incompleteness errors. The incorporation of F12 terms addresses this deficiency, yielding improved convergence with respect to basis size. This work demonstrates that explicitly correlated corrections are not limited to single-reference orbital frameworks, but can be naturally integrated with geminal-based theories, providing a unified route to treating both static and dynamic correlation efficiently.

\subsection{Transcorrelated Methods/Gaussian Geminals}

An alternative route to incorporating explicit electron–electron correlation is provided by the transcorrelated (TC) method, in which correlation effects are introduced through a similarity transformation of the Hamiltonian rather than by augmenting the wavefunction with explicit geminal terms. In this framework, the correlated wavefunction is written as $\Psi = e^\mathfrak{T} \Phi$, where $\mathfrak{T}$ is a correlation factor (typically a sum of two-electron functions, $f(r_{ij})$), and $\Phi$ is a Slater determinant. This leads tot the transcorrelated Hamiltonian, $\bar{H} = e^\mathfrak{-T}He^{\mathfrak{T}}$. which is used to solve a transformed Schr\"odinger equation.

A key advantage of this formulation, originally introduced by Boys and Handy, is that the transformed Hamiltonian terminates at the three-body operator level, even when the correlation factor contains pairwise terms. However, this comes at the cost of introducing non-Hermiticity and effective many-body interactions.

A practical realization of this idea was developed by Ten-no, who proposed a feasible transcorrelated approach using a frozen Gaussian geminal (FROGG)\cite{ten2000feasible}. n this formulation, the correlation factor is chosen as a sum of short-range Gaussian geminals,

\begin{equation}
    \mathfrak{T} \equiv F = \frac{1}{2} \sum_{i\neq j} f(r_{ij}) \quad f(r_{12}) = -\sum_G c_G \ e^{-\zeta_Gr_{12}^2}
\end{equation}

which is designed to cancel the Coulomb singularity at short interelectronic distances while remaining computationally manageable. The resulting similarity-transformed Hamiltonian takes the form
\begin{equation}
    \bar{H} = H + K +L,
\end{equation}

where $K$ and $L$ represent effective two- and three-body operators arising from the transformation. By employing a short-range geminal, the method ensures that the additional interactions are localized, leading to favorable scaling and enabling applications within perturbative frameworks such as MP2. This work demonstrated that the transcorrelated approach can significantly accelerate basis-set convergence while maintaining computational feasibility.

More recently, Lee and Thom revisited the transcorrelated framework and developed a self-consistent, bi-variational formulation of the TC method . In their approach, the effective Hamiltonian is constructed using a Jastrow-type correlator, 
\begin{equation}
    \mathfrak{T} = \sum_{i<j} u(\textbf{r}_i, \textbf{r}_j),
\end{equation}

and the transformed Hamiltonian is expressed via a Baker–Campbell–Hausdorff expansion, leading to effective one-, two-, and three-body terms. A central challenge of the TC method—the non-Hermiticity of $\bar{H}$ is addressed by employing a bi-orthogonal formulation, in which left and right eigenfunctions are treated independently. This allows for a consistent definition of the energy, $E = \langle \Psi_L | \bar{H} | \Psi_R \rangle$, and enables stable optimization of the wavefunction.

In addition, Lee and Thom introduced a second-order moment (SOM) minimization strategy to optimize the correlator parameters, providing an alternative to traditional variance minimization for non-Hermitian operators. Their results demonstrate that the transcorrelated method can achieve highly accurate energies and improved basis-set convergence for atoms and small molecules, while offering a deterministic alternative to stochastic approaches.

Recent developments have significantly expanded the applicability of TC methods by integrating them with advanced electronic structure frameworks and improving their computational scalability. A key direction has been the combination of TC Hamiltonians with stochastic and many-body approaches. Luo and Alavi demonstrated that the TC Hamiltonian can be efficiently combined with full configuration interaction quantum Monte Carlo (FCIQMC), where the projection-based nature of the method circumvents difficulties associated with the non-Hermiticity of the transformed Hamiltonian\cite{luo2018combining}. In this framework, the inclusion of explicit correlation through the TC transformation substantially accelerates basis-set convergence, improving the asymptotic scaling of the correlation energy with respect to the basis size. This work highlights that TC methods are particularly well suited for stochastic solvers, where the compactness of the transformed wavefunction translates directly into computational efficiency.

Beyond methodological combinations, recent studies have demonstrated the robustness of TC approaches across a wider range of chemical systems. In particular, Filip et al. extended TC methods to second-row elements, showing that chemically accurate energies and ionization potentials can be obtained with relatively modest basis sets\cite{filip2025transcorrelated}. These results confirm that the improved convergence properties observed in model systems and first-row atoms persist in heavier elements, and further emphasize that the TC Hamiltonian leads to a more compact representation of the correlated wavefunction, reducing the multiconfigurational character required for accurate descriptions.

Another important line of development focuses on improving the efficiency and scalability of TC methods for larger systems. Simula et al. introduced a formulation of transcorrelated theory incorporating pseudopotentials, which reduces the number of explicitly treated electrons and simplifies the optimization of the Jastrow correlator\cite{simula2025transcorrelated}. By eliminating electron–nucleus cusps and lowering the variance in variational Monte Carlo optimization, this approach significantly decreases computational cost while retaining high accuracy for molecular properties such as ionization potentials and dissociation energies. These advances open the door to applying TC methods to more complex systems, including transition-metal compounds and extended materials.

Finally, hybrid approaches that combine transcorrelation with explicitly correlated F12 techniques have emerged as a promising route to further improve accuracy and balance in electronic structure calculations. Masteran et al. developed a transcorrelated F12 framework in which geminal-based correlation factors are incorporated within a (unitary) similarity-transformed Hamiltonian\cite{masteran2025toward}. This approach retains the favorable basis-set convergence of F12 methods while benefiting from the compactness of the transcorrelated Hamiltonian, and enables a more balanced description of ground and excited states. By reducing basis-set requirements for both energies and response properties, such hybrid TC–F12 formulations illustrate a unification of wavefunction-based and Hamiltonian-based explicit correlation strategies.

Overall, transcorrelated methods provide a conceptually distinct yet closely related framework to F12 theories: both incorporate explicit dependence on the interelectronic distance to satisfy cusp conditions, but differ fundamentally in their implementation. While F12 methods augment the wavefunction with geminal terms, the transcorrelated approach instead renormalizes the Hamiltonian, embedding correlation effects directly into the operator. This distinction leads to different computational trade-offs—most notably the emergence of non-Hermiticity and three-body interactions in TC methods—but also offers a flexible and systematically improvable route to treating electron correlation.

\subsection{Jastrow–Geminal Methods}

A major conceptual advance in geminal theory emerged from its integration with Jastrow correlation factors within quantum Monte Carlo (QMC) frameworks. Casula and Sorella introduced the AGP–Jastrow (\acrshort{jagp}) ansatz\cite{casula2003geminal}, in which the many-electron wavefunction is written as
\begin{equation}
\Psi(\mathbf{r}_1,\ldots,\mathbf{r}_N)
= e^{J(\mathbf{r}_1,\ldots,\mathbf{r}_N)} \, \Psi_{\text{AGP}}(\mathbf{r}_1,\ldots,\mathbf{r}_N),
\end{equation}
where the antisymmetrized geminal power $\Psi_{\text{AGP}}$ captures nondynamic (static) correlation through electron pairing, while the symmetric Jastrow factor introduces explicit dynamical correlation. 

The Jastrow exponent is typically expressed as a sum of two-electron terms,
\begin{equation}
J(\mathbf{r}_1,\ldots,\mathbf{r}_N)
= \sum_{i<j} f(\mathbf{r}_i,\mathbf{r}_j),
\end{equation}
leading to the multiplicative form
\begin{equation}
e^{J} = \prod_{i<j} \exp\!\big(f(\mathbf{r}_i,\mathbf{r}_j)\big),
\end{equation}
which enforces the electron–electron cusp condition and accelerates convergence of the wavefunction. This construction yields a clear separation of roles: the geminal component describes static pairing correlations, while the Jastrow factor accounts for short-range dynamical correlation.

A key practical advantage of the JAGP ansatz is that the antisymmetric geminal component can be evaluated as a single determinant even for spin-polarized systems, leading to a computational cost comparable to Hartree–Fock despite its multiconfigurational character. In benchmark applications to atoms up to $Z=14$, the method recovered approximately 60–70\% of the correlation energy at the variational level and more than 90\% within fixed-node diffusion Monte Carlo, with particularly strong performance for multiconfigurational systems such as Be. Importantly, when optimized together with the geminal, the Jastrow factor also contributes to an indirect improvement of the nodal surface, enabling high-quality fixed-node energies without large multideterminant expansions. 

Beyond atomic systems, the JAGP framework naturally connects to the resonating valence bond (RVB) picture, in which electron pairing generates a superposition of valence-bond structures. In this interpretation, the geminal expansion provides a compact representation of an exponentially large configuration space, while the Jastrow factor suppresses unphysical configurations and enforces physically correct short-range behavior. While the original formulation emphasized the balanced treatment of static and dynamic correlation, subsequent studies clarified that the effectiveness of the JAGP ansatz in projection QMC is primarily governed by the quality of its nodal surface. This aspect was highlighted in all-electron quantum Monte Carlo calculations of the sodium dimer by Nakano \textit{et al.}\cite{nakano2019all}, where a JAGP guiding wavefunction within lattice-regularized diffusion Monte Carlo (LRDMC) achieved chemical accuracy for the binding energy, equilibrium distance, and vibrational frequency. Comparison with single-determinant Jastrow–Slater references revealed that the multiconfigurational nature of JAGP significantly improves the nodal structure, thereby reducing fixed-node errors and enhancing error cancellation between atoms and molecules. Notably, the improvement is more pronounced for the bonded system than for the isolated atoms, underscoring the importance of geminal-based static correlation in weak chemical bonding.Overall, the JAGP ansatz provides a unifying bridge between geminal theories, explicit correlation, and modern QMC methods, demonstrating that accurate many-body descriptions can be achieved using compact, determinant-like wavefunctions augmented by physically motivated correlation factors.

While the original formulation emphasized the balanced treatment of static and dynamic correlation, subsequent studies clarified that the effectiveness of the JAGP ansatz in projection QMC is primarily governed by the quality of its nodal surface. This aspect was highlighted in all-electron quantum Monte Carlo calculations of the sodium dimer by Nakano \textit{et al.}\cite{nakano2019all}, where a JAGP guiding wavefunction within lattice-regularized diffusion Monte Carlo (LRDMC) achieved chemical accuracy for the binding energy, equilibrium distance, and vibrational frequency. Comparison with single-determinant Jastrow–Slater references revealed that the multiconfigurational nature of JAGP significantly improves the nodal structure, thereby reducing fixed-node errors and enhancing error cancellation between atoms and molecules. Notably, the improvement is more pronounced for the bonded system than for the isolated atoms, underscoring the importance of geminal-based static correlation in weak chemical bonding.

Building on this understanding, large-scale benchmarks have demonstrated that improved nodal surfaces translate directly into chemical accuracy across diverse systems. Raghav et al. employed a Jastrow-correlated antisymmetrized geminal power with singlet correlation (JsAGPs) ansatz within variational and LRDMC calculations to compute atomization energies for the 55-molecule Gaussian-2 (G2) test set\cite{raghav2023toward}.
The resulting wavefunction takes the form,
\begin{equation}
    \Psi_{\mathrm{JsAGPs}}(\mathbf{R}) = e^{J(\mathbf{R})} \det \big[f(\mathbf{r}_i,\mathbf{r}_j)\big],
\end{equation}
where $J(\mathbf{R})$ is a many-body Jastrow factor accounting for electron–electron and electron– nucleus correlations, and $f(\mathbf{r}_i,\mathbf{r}_j)$ is a geminal (pairing) function expanded in an atomic orbital basis.

This formulation retains a compact single-determinant form while implicitly incorporating multideterminant character through the geminal expansion.
  
By variationally optimizing both the Jastrow and geminal parameters, including the nodal surface, at the variational Monte Carlo level, followed by LRDMC projection, the authors achieved a mean absolute deviation of 1.6 kcal/mol for atomization energies, compared to 3.2 kcal/mol for Jastrow–DFT determinant references, reaching chemical accuracy for a substantial portion of the dataset. These results demonstrate that the nodal improvements afforded by geminal-based pairing translate into systematically enhanced predictive accuracy, positioning JAGP-based approaches as a promising intermediate between single-determinant and fully multideterminant methods for high-accuracy electronic structure calculations.
 

A fundamental limitation of the antisymmetrized geminal power (AGP) ansatz is its lack of size consistency, arising from unphysical charge fluctuations between noninteracting subsystems. Neuscamman demonstrated that this deficiency can be completely removed by introducing a network of location-specific Hilbert-space Jastrow factors that enforce local particle-number constraints.\cite{neuscamman2012size} In this formulation, the Jastrow operator is written as
\begin{equation}
\hat{J} = \sum_{pq} \sum_{n,m \in {0,1}} C^{pq}_{nm}, \hat{P}^p_n \hat{P}^q_m,
\end{equation}
where $\hat{P}^p_n$ projects onto occupation $n$ of orbital $p$ and $C^{pq}_{nm}$ are variational parameters. The term “location-specific” reflects that these Jastrow factors act on selected orbitals or spatial regions (e.g., atoms or fragments), allowing the wavefunction to penalize or favor particular local occupation patterns. In effect, this Jastrow network acts as a flexible particle-number projector that suppresses charge-transfer (“ionic”) configurations responsible for size inconsistency. The origin of the AGP size-consistency error can be traced to such unphysical ionic terms, in which multiple electron pairs occupy the same local geminal, thereby preventing proper factorization across noninteracting subsystems. By enforcing local electron-number constraints, the Jastrow network restores the correct separability of the wavefunction while retaining polynomial computational scaling and a fully variational framework.

Building on this foundation, the Jastrow–antisymmetrized geminal power (JAGP) ansatz was subsequently developed as a practical electronic structure method in Hilbert space.\cite{neuscamman2013jastrow} By combining a nonorthogonal AGP reference with a Jastrow factor, the ansatz captures both the resonating valence-bond character of electron pairing and interpair correlations absent in traditional geminal theories. Efficient determinant-based evaluation techniques enable variational optimization with polynomial $O(N^5)$ scaling. Applications to molecular bond dissociation and transition states demonstrate that JAGP achieves near–active-space accuracy for static correlation while maintaining a compact wavefunction form.

A further conceptual advance was introduced through the cluster Jastrow antisymmetrized geminal power (CJAGP) framework, which reinterprets the role of the Jastrow factor in a fundamentally subtractive manner.\cite{neuscamman2016subtractive} In this formulation, the Hilbert-space Jastrow operator,

\begin{equation}\label{eq:jastrow}
\hat{J} = \sum_{pq} J_{pq}, \hat{n}_p \hat{n}_q,
\end{equation}
where $\hat{n}$ represent the number operators. This operator defines a transformation $\hat{Q} = e^{\hat{J}}$ that can also be viewed as a constrained doubles cluster operator whose amplitudes are determined by the Jastrow parameters. Rather than incrementally improving a physically motivated reference, the AGP component is allowed to generate an overcomplete and even unphysical set of configurations, while the Jastrow operator selectively suppresses undesirable ionic contributions. This “subtractive” mechanism enables highly accurate wavefunctions despite the poor standalone quality of the geminal reference, drawing a close analogy to variation-after-projection approaches.

Subsequent developments established the practical robustness of the CJAGP ansatz through an improved variational Monte Carlo optimization scheme based on the linear method.\cite{neuscamman2016improved} By reformulating the stochastic evaluation of the linear-method matrices, the computational scaling was reduced to $O(N^5)$ per sample, with further reduction to $O(N^4)$ using Krylov-subspace techniques, while retaining a near zero-variance property that significantly enhances statistical efficiency. This approach enables stable optimization of all wavefunction parameters, including orbitals, and yields accurate potential energy surfaces for strongly correlated systems such as N$_2$ and [ScO]$^+$, outperforming conventional single-reference methods and competing with multireference approaches at polynomial cost.

\noindent Khamoshi \textit{et al.} investigated non-linear exponential correlators built on the AGP reference, extending earlier linear geminal replacement approaches\cite{khamoshi2021exploring}. They studied two classes of correlators. First, Hilbert-space Jastrow operators of the form presented in the eq.(\ref{eq:jastrow}), were treated through a similarity-transformed Hamiltonian $\bar{H} = e^{-\hat{J}} H e^{\hat{J}}$, which can be summed analytically to all orders within the pairing algebra. While this construction enables a CC-like formalism with polynomial scaling when combined with AGP reconstruction formulas, numerical results show that the resulting Jastrow-based exponential ansatz can exhibit overcorrelation and does not systematically outperform linear correlator methods. 

In contrast, the unitary pair-hopper Ansatz,
\begin{equation}
    |\Psi\rangle = e^{\hat{\mathcal{T}}} |\mathrm{AGP}\rangle, \qquad
\mathcal{T} = \sum_{p<q} \tau_{pq} \big( P_p^\dagger P_q - P_q^\dagger P_p \big),
\end{equation}
provides a more effective non-linear exntension by generating correlated pair excitation to unitary CC theory. Approximations based on truncated Baker–Campbell–Hausdorff expansions and canonical transformation theory yield accurate energies across weak and strong correlation regimes, often improving upon linear geminal CI approaches. These results highlight that the success of exponential post-AGP theories depends critically on the structure of the correlator: while Jastrow exponentials act primarily as multiplicative reweighting of configurations, excitation-based operators such as the pair-hopper more closely reproduce the favorable behavior of coupled-cluster methods. Consequently, AGP emerges not merely as a static reference, but as a flexible platform for constructing diverse non-linear correlator theories, whose effectiveness is strongly ansatz-dependent.


\subsection{Geminal Replacement and J-CI}

A unifying perspective on post-AGP correlation was provided by Dutta, Henderson, and Scuseria, who demonstrated that seemingly distinct AGP-based correlator constructions—killer operators, number-operator (Hilbert-space Jastrow) correlators, and pair-hopping excitations—are algebraically equivalent representations of geminal replacement processes.\cite{dutta2020geminal} In this framework, correlation on AGP is interpreted as the systematic replacement of one or more geminals, in direct analogy to orbital replacement in configuration interaction. The resulting J$_k$-CI hierarchy, formulated in terms of number-operator correlators, yields a sequence of systematically improvable seniority-conserving models. Numerical studies on the pairing Hamiltonian show that J$_2$-CI substantially outperforms conventional CID and CCD in the strong-correlation regime, while higher-order J$_k$-CI approaches the full seniority-zero (DOCI) limit. Moreover, symmetric tensor decomposition of the correlator amplitudes reveals that higher-order J-CI naturally generates linear combinations of AGPs (LC-AGP), establishing a formal bridge between AGP and the more general APIG ansatz. This work recasts post-AGP correlation as a structured geminal-replacement hierarchy and clarifies the algebraic relationships among previously proposed correlator forms.


\section{Alternative Theoretical Perspective}


\subsection{2RDM/statistical geminal energy reconstruction}

An alternative perspective on geminals emerges from reduced-density-matrix (RDM) mechanics, where geminals appear as eigenfunctions of the two-particle reduced Hamiltonian $\hat{K}^{(2)}$. Rothman and Mazziotti reformulated the many-electron ground-state energy as a linear functional of the eigengeminals of $\hat{K}^{(2)}$.
\begin{equation}
    E = \sum_n p_n \varepsilon_n
\end{equation}

where $\varepsilon_n$ are the eigenvalues of $\hat{K}^{(2)}$ and $p_n$ are their occupation numbers\cite{rothman2008geminal}. Rather than constructing an explicit geminal wavefunction, the method assigns occupations to these “pairon” states through a statistical model inspired by Fermi–Dirac distributions, introducing a correlation temperature that parametrizes the redistribution of pair populations. In the zero-temperature limit, the scheme reproduces the Hartree–Fock energy, while finite correlation temperatures incorporate correlation effects through the correlated two-electron spectrum. This approach reframes the electronic structure problem as one of statistically populating correlated two-electron eigenmodes, offering a computationally simple but non-variational alternative to constrained 2RDM optimization. Although the predictive protocol relies on empirical calibration of the correlation temperature, the formulation highlights a complementary viewpoint: correlation may be understood either through variational pair amplitudes in a wavefunction ansatz or through occupation statistics of correlated geminal eigenstates in reduced-density-matrix space.

Analysis of accurate multi-reference configuration interaction two-particle density matrices has revealed that a geminal-like structure emerges naturally in the natural orbital basis, even when no explicit geminal wavefunction is assumed. These studies show that while APSG-type models capture intrageminal correlation, significant dynamical correlation arises from intergeminal multipole–multipole interactions that are absent in simple JKL-type density-matrix functionals. \cite{van2018non}

A recent development in reduced density matrix theory combines geminal-based representations with machine learning to address the long-standing challenge of determining N-representable 2-RDMs. In this approach, the electronic energy is expressed as a functional of the two-electron reduced density matrix in its diagonal geminal basis, where it reduces to a weighted sum of geminal energies. While the geminal eigenvalues are readily obtained from a two-electron Hamiltonian, the corresponding occupation numbers remain difficult to determine due to N-representability constraints. Sager-Smith and Mazziotti introduced a machine learning framework in which these occupations are modeled through a Boltzmann-like distribution parameterized by an effective correlation temperature, learned via a neural network\cite{sager2022reducing}. This formulation enables the reduction of the many-electron problem to an effective two-electron problem while implicitly enforcing N-representability, thereby providing a novel data-driven realization of 2-RDM theory with favorable computational scaling.
\section{Specialized Extensions}

\subsection{Geminals for Open-Shell and Odd-Electron Systems}

While geminal wavefunctions are naturally tailored to systems with an even number of electrons, strategies have been developed to extend them to open-shell or odd-electron cases. One approach augments the antisymmetrized product of geminals (APG) with an additional single-particle operator, while another relies on a more general characterization of the pair-vacuum state. For recent advances on extending geminal-based formalisms beyond the closed-shell setting, see Johnson et al. \cite{johnson2017strategies}.


A notable development in the geminal landscape has been the introduction of Pfaffian-based ansätze in quantum Monte Carlo (QMC) simulations. By replacing determinants with Pfaffians, researchers were able to antisymmetrize general fermionic pairing functions that include both singlet and triplet channels, and augment them with a flexible many-body Jastrow factor. This Jastrow–Pfaffian framework (JAGP) \cite{genovese2020general} dramatically improves the description of electronic correlation while retaining computational costs comparable to single-determinant wavefunctions. Benchmark studies on notoriously challenging systems such as C$_2$, N$_2$, O$_2$, and benzene demonstrate accuracy rivaling or surpassing state-of-the-art quantum chemistry methods, including CCSD(T) and DMRG, even with compact basis sets. The inclusion of triplet correlations proved essential for correctly capturing spin fluctuations and binding in open-shell and strongly correlated dimers, making JAGP a size-consistent, chemically accurate, and computationally viable ansatz. These results underscore the versatility of geminal-inspired forms when combined with stochastic optimization, and suggest that Pfaffian-based extensions could become a powerful paradigm for scalable, accurate electronic structure modeling.

Bajdich et al. introduced Pfaffian pairing wavefunctions as a powerful generalization of determinant- and BCS-based ansätze for quantum Monte Carlo (QMC), enabling a unified treatment of singlet, triplet, and unpaired electrons within a single antisymmetric form\cite{bajdich2006pfaffian}. In contrast to Slater determinants, which encode antisymmetry through one-particle orbitals, Pfaffian wavefunctions are built from pair (geminal) orbitals, thereby naturally incorporating electron pairing correlations. A key contribution of this work is the formulation of the singlet–triplet–unpaired (STU) Pfaffian, which combines all pairing channels in a compact algebraic structure,

\begin{equation}
    \Psi_{STU} = \text{Pf} \begin{pmatrix}
        \mathbf{\xi}^{\uparrow \uparrow} & \mathbf{\phi}^{\uparrow \downarrow} & \mathbf{\varphi}^\uparrow \\ 
        - \mathbf{\phi}^{\uparrow \downarrow T} & \mathbf{\xi}^{\downarrow \downarrow} & \mathbf{\varphi}^\downarrow \\
        -\mathbf{\varphi}^{\uparrow T} & -\mathbf{\varphi}^{\downarrow T}  & 0
    \end{pmatrix},
\end{equation}

where $\mathbf{\phi, \xi}$, and $\mathbf{\varphi}$ denote singlet pairing, triplet pairing, and unpaired orbitals, respectively\cite{bajdich2006pfaffian}.his unified representation recovers Hartree–Fock and BCS wavefunctions as limiting cases, while providing a systematically improvable framework through multipfaffian expansions. Importantly, Pfaffian wavefunctions yield substantially improved fermionic nodal structures, reducing the artificial nodal partitioning of Hartree–Fock to the physically correct minimal topology, which directly enhances the accuracy of fixed-node diffusion Monte Carlo. Benchmark calculations on first-row atoms and molecules demonstrate that compact Pfaffian forms recover a large fraction ($\approx$95–99\%) of the correlation energy, with further improvements achievable via multipfaffian and backflow extensions. Overall, Pfaffian-based geminal wavefunctions offer a physically transparent and highly flexible route to capturing both static and dynamic correlation, particularly through their superior description of fermionic nodes.

Recent developments have extended geminal-based methods to systems with strong spin correlations by relaxing the conventional singlet-pairing constraint. Szabados, Mihálka, and Surján introduced a spin-unrestricted antisymmetrized product of strongly orthogonal geminals (APSG) framework combined with spin projection and orbital optimization, enabling a balanced description of open-shell and biradical systems\cite{szabados2025spin}. In this approach, the reference wavefunction is constructed as a product of geminal creation operators, as shown in eq.(\ref{eq:general_geminal}), with each geminal expanded in one-particle orbitals. 

A key point of this formulation is that the coefficient matrix $C$ has both symmetric and asymmetric components. This naturally leads to a decomposition of each geminal into singlet and triplet pairing contributions,

\begin{equation}
    C_{ij} = C_{ij}^{(S)} + C_{ij}^{(T)}, \quad C_{ij}^{(S)} = \frac{1}{2} (C_{ij} + C_{ji}), \quad C_{ij}^{(T)} = \frac{1}{2} (C_{ij} - C_{ji}),
\end{equation}
which gives rise to an intrinsic mixture of singlet and triplet pairing within each geminal. In contrast to conventional APSG models that enforce pure singlet pairing, this spin-unrestricted formulation allows the wavefunction to adapt flexibly to near-degeneracy and spin recoupling effects.

The resulting spin contamination is subsequently mitigated through the application of a projection operator, such as the half-projection operator,

\begin{equation}
    \hat{P}_{\text{HP}} = \frac{1}{2} \left(1 + f \hat{P}\right)
\end{equation}

which selectively restores the desired spin symmetry while retaining the benefits of symmetry breaking. A key innovation of this work is the fully variational optimization of both geminal coefficients and orbitals, combined with a symmetry-adapted perturbation theory (PT) correction to recover dynamic correlation. Applications to cyclobutadiene demonstrate that this framework captures the delicate balance between singlet and triplet states, providing a qualitatively correct description of the automerization barrier and singlet–triplet gap. The study highlights that controlled spin-symmetry breaking and restoration are essential for extending geminal methods to strongly correlated open-shell systems, offering an economical alternative to multireference approaches


A perturbative extension of \acrfull{slg} based on the unrestricted Hartree–Fock (UHF) wavefunction provided a pivotal bridge between mean-field symmetry breaking and geminal model chemistry.\cite{foldvari2019geminal} By exploiting the natural orbital representation of UHF, the reference was reinterpreted as a product of strongly orthogonal geminals containing both singlet and triplet components. A Dyall-type zero-order Hamiltonian was constructed in which intrageminal electron–electron interaction is treated explicitly, while intergeminal correlation is incorporated perturbatively. The resulting UNO–SLG perturbation theory (UNOSLGPT) avoids the severe imbalance characteristic of UMP for spin-contaminated references and restores reliable singlet–triplet splittings in biradical systems such as H$_4$ and O$_2$. Formal analysis clarified how spin contamination arises at the intergeminal level and how relaxation of geminal coefficients at order zero can partially mitigate it. This work established UHF-derived geminals as minimal static-correlation references and laid the groundwork for subsequent developments involving spin projection and symmetry-adapted perturbation corrections.

Half-projection of strongly orthogonal unrestricted geminal products (HP-SLG) was introduced as an intermediate alternative between spin-contaminated but size-consistent unrestricted geminals and fully spin-projected, size-inconsistent schemes\cite{mihalka2019half}. In this construction, a simplified projection operator is applied to an antisymmetrized product of strongly orthogonal geminals, eliminating terms with an odd number of triplet components while preserving computational economy. Variation after half-projection leads to effective two-electron eigenvalue equations for each geminal, retaining the favorable mean-field-like structure of the original SLG model. Numerical tests on H$_4$ and biradical systems demonstrated substantial reduction of spin contamination relative to the unprojected reference, while the associated size-consistency error was significantly smaller than that of fully projected schemes. Although size consistency is not fully restored, the violation is intergeminal in origin and remains moderate for chemically relevant systems, establishing HP-SLG as a practical compromise between symmetry restoration and extensivity.

A symmetry-adapted perturbative correction to half-projected strictly localized geminals (HP-SLG) was subsequently developed within a SAPT-inspired formalism.\cite{mihalka2021symmetry} In this approach, the unprojected geminal product defines the zero-order Hamiltonian, while spin purification is incorporated at the perturbative level via weak or strong symmetry forcing. This construction avoids the severe imbalance observed in perturbation theories built on spin-contaminated references and restores reliable singlet–triplet splittings in biradicaloids, where unprojected PT fails. Careful partitioning of the effective geminal Hamiltonian was shown to be essential to prevent artificial intruder states and to maintain the correct spin-pure limit. These developments established half-projected geminals as viable multireference starting points for perturbative dynamic correlation.

Building on this foundation, full variational optimization of half-projected antisymmetrized products of strongly orthogonal geminals (HP-APSG) has recently been implemented.\cite{szabados2025orbital} In this formulation, the half-projection operator is applied prior to variation, suppressing Coulson–Fischer–type instabilities while retaining the flexibility of unrestricted pairing. Both geminal coefficients and molecular orbitals are optimized self-consistently, and dynamical correlation is incorporated via symmetry-adapted perturbation theory. Applications to cyclobutadiene along its automerization coordinate show that strong symmetry forcing within the perturbative correction yields quantitatively accurate singlet–triplet gaps, whereas orbital optimization itself plays a comparatively minor energetic role. Together, these developments consolidate spin unrestriction, symmetry restoration, and perturbative corrections into a unified strongly orthogonal geminal framework suitable for biradical and near-degenerate systems.


\subsection{Geminals in the Quantum Computing Era}

Geminal-based ans\"atze have recently gained significant attention in the context of quantum algorithms for electronic structure. This interest is largely driven by the limitations of current quantum hardware, often referred to as \emph{noisy intermediate-scale quantum} (NISQ) devices, which are characterized by a limited number of qubits, short coherence times (the duration a qubit can maintain its fragile quantum state such as superposition or entanglement, before losing information to the environment), and the absence of full error correction. In this regime, it is essential to employ wavefunction representations that are both physically expressive and resource-efficient. The intrinsic orbital-pairing structure of geminal wavefunctions makes them particularly attractive candidates for such applications.

A major line of development in this direction has been driven by Khamoshi, and co-workers, who systematically established the role of antisymmetrized geminal power (AGP) wavefunctions in quantum computing through a sequence of complementary advances.

In their initial work, Khamoshi \textit{et al.} demonstrated that AGP can serve as an efficient and physically meaningful reference for quantum algorithms\cite{khamoshi2020correlating}. They showed that AGP can be implemented on a quantum computer with circuit depth and gate count scaling linearly with system size, by exploiting its equivalence to a number-projected BCS state and mapping fermion pairs directly to qubits. Furthermore, they introduced unitary correlators acting on AGP—constructed from its associated “killing operators”—and showed that this framework yields highly accurate energies across both weakly and strongly correlated regimes, particularly for pairing Hamiltonians.

Building on this foundation, subsequent work formulated a systematic extension of unitary coupled-cluster theory using AGP as the reference state\cite{khamoshi2022agp}. In this approach, excitation operators are derived from the AGP killing operators, which annihilate the reference and define a compact, physically adapted excitation manifold. This leads to a disentangled unitary coupled-cluster ansatz that preserves the pairing structure of the wavefunction while incorporating correlations beyond AGP. Compared to conventional UCCSD based on Hartree–Fock, the AGP-based construction avoids redundant excitations and provides a more balanced description of strong correlation, establishing a direct conceptual bridge between geminal theory and unitary coupled-cluster frameworks.

A key practical challenge in this work, namely, the efficient preparation of AGP states on quantum hardware—was addressed in later work\cite{khamoshi2023state}. By exploiting the equivalence between AGP and elementary symmetric polynomial states, the authors developed a deterministic and polynomial-cost state preparation algorithm that avoids explicit number projection. This construction enables shallow quantum circuits and provides a hardware-friendly realization of AGP, completing the framework needed to deploy geminal-based ans\"atze within variational quantum eigensolver algorithms.

These contributions establish a unified paradigm in which AGP serves simultaneously as a compact reference state, a foundation for constructing tailored unitary correlators, and a practical ansatz realizable on near-term quantum devices. This progression highlights how geminal-based ideas can be systematically adapted into resource-efficient quantum algorithms.

Complementary to these approaches, efforts were taken in proposing the resource-efficient quantum algorithms for seniority-zero subspaces, where only doubly occupied configurations are retained. In this context, Elfving \textit{et al.} proposed a paired-electron unitary coupled-cluster doubles (pUCCD) ansatz\cite{elfving2021simulating}, which is closely related to DOCI and AP1roG. By restricting the Hilbert space, the method achieves linear circuit depth and quadratic measurement scaling, while still capturing essential static correlation effects. For a fixed number of qubits, this reduction in complexity can be traded for larger basis sets, often improving overall accuracy despite the imposed constraint.

Beyond pairing-based truncations, recent work has explored the incorporation of Jastrow- and geminal-inspired factorizations directly at the level of the quantum circuit ansatz. In this vein, Matsuzawa and Kurashige introduced a unitary cluster Jastrow (uCJ) decomposition, in which generalized coupled-cluster doubles amplitudes are factorized into products of Jastrow-type operators, leading to the $k$-uCJ ansatz with only $\mathcal{O}(N^2)$ parameters. Formally, this corresponds to expressing the cluster operator as a sequence of unitary transformations of the form
\begin{equation}
    \hat{U}_{\text{uCJ}} = \prod_{x=1}^k e^{\hat{K}^{(x)}} e^{\hat{J}^{(x)}} e^{-\hat{K}^{(x)}},
\end{equation}
where $\hat{K}$ generates orbital rotations and $\hat{J}$ is a Jastrow-type operator composed of number operators. Bthe key advantage of this construction is that the $\hat{J}$ operators are built from commuting number operators, which eliminates the need for Trotter decomposition when implementing the ansatz on quantum circuits\cite{matsuzawa2020jastrow}. This leads to significantly reduced circuit depth compared to conventional unitary coupled-cluster approaches. 

Another direction of interest came forward in using explicitly correlated and transcorrelated methods as tools for reducing the computational complexity of electronic structure calculations on near-term quantum hardware. A central challenge in quantum algorithms for chemistry is the large number of qubits and circuit depth required to accurately represent electron correlation using conventional orbital-based Hamiltonians. In this context, TC methods offer a compelling advantage by incorporating short-range correlation effects directly into an effective Hamiltonian, thereby yielding more compact wavefunctions.


McArdle and Tew demonstrated that the transcorrelated Hamiltonian can significantly reduce the quantum resources required for electronic structure calculations\cite{mcardle2020improving}. By applying a similarity transformation using a Jastrow-type correlator, the TC Hamiltonian removes the electron–electron cusp and redistributes correlation effects into modified one-, two-, and three-body terms. Importantly, this leads to wavefunctions that are closer to single-reference character and therefore require fewer excitations to represent on a quantum computer. To address the non-Hermiticity of the TC Hamiltonian, the authors employed an imaginary-time evolution framework, enabling stable computation of ground-state energies within a quantum algorithmic setting.

Building on this idea, Dobrautz et al. demonstrated that transcorrelated methods can be used to achieve near chemical accuracy on current quantum hardware by reducing both qubit requirements and circuit depth\cite{dobrautz2024toward}. In their approach, the TC transformation is combined with downfolding techniques to construct an effective Hamiltonian in a reduced orbital space, which can then be mapped efficiently onto qubits. The resulting framework enables accurate simulations of molecular systems using significantly fewer quantum resources than required by conventional formulations. Notably, this work highlights that incorporating explicit correlation at the Hamiltonian level can be more advantageous than attempting to recover it variationally within a quantum circuit.

A more recent and significant advance in this direction was reported by Sokolov et al., who developed an exact transcorrelated framework tailored for quantum computing applications\cite{sokolov2023orders}. In contrast to earlier approximations that simplify the transformed Hamiltonian, their approach retains the full non-Hermitian structure arising from the similarity transformation and addresses the associated challenges using a variational quantum imaginary-time evolution (VarQITE) algorithm. A key advantage of this formulation is the dramatic compactification of the ground-state wavefunction, which becomes nearly single-reference in character. As a result, accurate solutions can be obtained using significantly shallower quantum circuits, even for strongly correlated systems such as the Hubbard model. Notably, their simulations and hardwarurce-constrained quantum algorithms. This emerging synergy between geminal theory and quantum computing suggests a promising direction for achieving chemically accurate simulations on near-te demonstrations show orders-of-magnitude improvements in both energy accuracy and wavefunction fidelity, highlighting the potential of transcorrelated methods to substantially reduce quantum resource requirements. This work establishes the transcorrelated approach as a powerful and practical framework for achieving high-accuracy quantum simulations on near-term devices.

Taken together, these developments illustrate a unifying theme: ideas originally developed in geminal and explicitly correlated theories—such as pairing structure, Jastrow factors, and cusp-corrected Hamiltonians—translate naturally into resource-efficient quantum algorithms. By embedding correlation effects either in the wavefunction ansatz or directly into the Hamiltonian, these approaches provide a promising pathway toward chemically accurate simulations on near-term quantum devices.


\section{Future Outlook}




Geminal wavefunctions, once considered computationally prohibitive, are reemerging as versatile tools for strongly correlated systems with efforts to incorporate missing dynamical correlations. Future opportunities include:
(a) Integrating geminal approaches with coupled-cluster and tensor-network formalisms to balance accuracy and efficiency.
(b) Extending geminal models for excited states, open-shell species, and dynamical correlation.
(c) Leveraging geminal-based ansätze for scalable quantum computing simulations of complex molecular systems.

The continued refinement of these methods reflects a broader trend: Pairing-based wavefunctions, rooted in early concepts of bonding and superconductivity, remain central to modern electronic structure theory. Their balance of chemical intuition, variational flexibility, and computational adaptability ensures that geminals will play a growing role in the future landscape of physical chemistry.

\section{Acknowledgment}
We acknowledge insights from timely discussions with Paul Ayers.
Also, we acknowledge support from the National Science Foundation CAREER award CHE-243986.









\printnoidxglossary[type=\acronymtype]

\bibliography{achemso-demo}

\end{document}